\documentclass[aps,physrev,reprint,groupedaddress]{revtex4-2}

\usepackage{graphicx}
\usepackage{amsfonts}
\usepackage{amsmath}
\begin{document}

%\preprint{}

\title{Approximating Grassmann valued path integrals with radial basis function neural networks}


\author{G\'abor Balassa}
%\thanks{}
\affiliation{Department of Physics, Yonsei University, Seoul 03722, Korea}
\affiliation{Institute for Particle and Nuclear Physics, HUN-REN Wigner Research Centre for Physics, 29-33 Konkoly-Thege Mikl\'os \'ut, Budapest, 1121, Hungary}
\email[]{balassa.gabor@yonsei.ac.kr}


\date{\today}

\begin{abstract}
Solving path integrals in quantum field theories often involves the numerical handling of noncommuting Grassmann fields, which is in many cases a highly nontrivial and numerically inefficient task, especially in large systems and at higher dimensions. In this paper a radial basis function type neural network construction is used to approximate fermionic path integrals that include local couplings in their hopping terms. By isolating the interaction terms from the purely fermionic components using a radial basis function expansion, the path integral can be approximated by a few percent accuracy even for very large lattice sizes. 
The method has been developed and tested using staggered fermions in one, and in two dimensions, through calculating the partition functions, and expectation values.
\end{abstract}

%\keywords{}

\maketitle


\section{Introduction}

Solving path integrals in quantum field theories is in general a challenging numerical task, especially when one wants to examine physically relevant observables that can be measured and be used to validate the underlying physical model \cite{1,2}. As in most of the cases that are physically interesting, analytical solutions do not exist, thus, one needs to rely on numerical methods, e.g., Monte Carlo sampling \cite{3}, tensor networks \cite{4}, complex Langevin dynamics \cite{5}, etc. One of the common properties of all of these methods is that they could take a very long time to be able to produce results with a satisfying accuracy. This problem is especially hard to overcome in gauge theories where the local gauge field is coupled to the derivative part of the fermionic action, which means the fermion determinant has to be calculated for each gauge configuration. For large systems (e.g., when taking the infinite volume limit), this is numerically infeasible. Partly due to this reason, lattice calculations, for example, in quantum chromodynamics \cite{6,7,8}, need access to high-computing facilities that allow fast parallel computation, and even then the calculations could take months to finish depending on the problem at hand.
Other than the time complexity issues, other numerical or conceptual problems could also arise. One of the most notorious is the so-called sign problem \cite{9,10}, which appears, e.g., in the description of the strong interaction at finite densities and restricts many methods that rely on importance sampling. 

There are an increasing number of attempts that can be found in the literature where different neural network configurations or machine learning techniques are applied to quantum mechanics, path integrals, or field theoretical problems (see e.g., \cite{16,N1,N2,N3,EE1,EE2,EE3,EE4,EE5}), where each method tries to tackle specific problems with a different approaches. In this work, we aim to give an alternative method for solving Euclidean path integrals of Grassmann fields by using carefully constructed radial basis function (RBF) type neural networks. 
In \cite{G1} the main idea of the method has been described and applied to interacting real scalar fields, where the phase transition line between the broken and unbroken phases was adequately described by expanding the quartic interaction terms using the RBF model. Moreover, in \cite{G2}, the model is used to overcome the sign problem in the context of interacting complex scalar fields at finite densities that suffer the same sign problem as quantum chromodynamics, thus it is an ideal test bed for methods that aim to solve the problem. With the RBF model, the corresponding phase transition and also the silver blaze phenomenon were adequately described in a very fast manner, where the whole calculation only took minutes on a standard notebook. By introducing noncommuting Grassmann fields \cite{11} into the picture, the RBF model had to be modified due to the fact that the training of the model needs ordinary numbers to be able to expand the interaction terms. It will be shown that by using the properties of the Grassmann variables and the structure of the RBF model, it is possible to extend the model to fermions as well. 

As this paper serves as a first attempt to include Grassmann valued fields into the RBF approximation, it will use relatively simple examples to show the working principles of the model. The extensions to more complex systems are briefly discussed, however problems in four dimensions, other fermion discretizations, or the inclusion of SU(N) gauge fields are left to future works.

First, in Sec.~\ref{sec:11} the general concept of the RBF expansion is introduced. The details of the model and the specific problem are then developed step by step through illustrative examples. The first example  in Sec.~\ref{sec:12} considers the one-dimensional path integral of staggered fermions with on-site interaction terms between the field $U_i$ and the fermions, such as $\bar{\chi}_i U_i \chi_{i+1}$, and $\bar{\chi}_i U_i \chi_{i-1}$. The second example in Sec.~\ref{sec:13} extends this framework to include interactions with neighboring sites, incorporating both $U_i$ and $U_{i-1}$, i.e., $\bar{\chi}_i  U_i \chi_{i+1}$ and $\bar{\chi}_i  U_{i-1} \chi_{i-1}$. Once the necessary theoretical and practical ingredients have been established, in Sec.~\ref{sec:2} the extension to higher dimensions will be given, while in Sec.~\ref{sec:3} the method to calculate expectation values within the RBF framework is discussed.

At the end, in Sec.~\ref{sec:4} some additional comments, and remarks will highlight the strengths and current limitations of the model, along with possible directions for extending the approach to more complicated theories.

\section{Radial basis function expansion for Grassmann variables}
This section provides an overview of the RBF expansion applied to Euclidean path integrals that include noncommuting Grassmann fields. It also introduces the parametrization and training methods required for the systems discussed later.
To illustrate the model’s working principles and establish the key components needed for the next section, the basic framework will be built up and validated through two simple examples using staggered fermions in one dimension.

\subsection{General description}
\label{sec:11}
In quantum field theories in the Euclidean formalism, one of the main objects of interest is the partition function that is a functional integral over all the possible field configurations weighted by the exponentialized action integral of the system. Schematically, a system with $\bar{\chi}(x)$ and $\chi(x)$ fermion degrees of freedom, where the action consists of interaction terms between the fermions and some external $U(x)$ fields, can be written as
\begin{equation}
\mathcal{Z} = \int \mathcal{D}\bar{\chi} \, \mathcal{D}\chi \, \mathcal{D}U \, \exp(-S[\bar{\chi},\chi,U]),
\end{equation}
where $\bar{\chi}$ and $\chi$ are Grassmann fields corresponding to the fermions, while $S[\bar{\chi},\chi,U]$ is the action integral of the underlying dynamical system. The RBF expansion will be applied to the discretized version of the path integral in accordance with \cite{G2}, however, as now the $\bar{\chi}$ and $\chi$ fields are Grassmann valued, it is not as straightforward to expand the interaction terms as it was the case for scalar fields. To make things more specific, let us write down the discretized version of the action at site $x$ for staggered fermions with a general gauge coupling in its hopping terms:
\begin{eqnarray}
\label{eq:2}
S_x[\bar{\chi},\chi,U_{\mu}] &=& a^4 \Bigg[ 
   \sum_{\mu} \frac{\eta_{\mu}(x)}{2a} 
   \Big( \bar{\chi}(x) U_{\mu}(x) \chi(x\!+\!\mu) 
\\
&& 
   - \bar{\chi}(x) U^{\dagger}_{\mu}(x\!-\!\mu) \chi(x\!-\!\mu) \Big) 
   + m \, \bar{\chi}(x)\chi(x) 
\Bigg] \nonumber,
\end{eqnarray}
where $\chi(x)$ is a one-component staggered fermion field at site $x$, $\bar{\chi(x)}$ is its conjugate field, $U_{\mu}(x)$ is the gauge link variable in direction $\mu$, $m$ is the mass of the fermion, $a$ is the lattice spacing, and $\eta_{\mu}(x)=(-1)^{\sum_{\nu<\mu}x_{\nu}}$ is the staggered phase factor \cite{12}. Using staggered fermions in the RBF method is advantageous because, unlike the naive Dirac fermions where one must work with four-component spinor fields, staggered fermions reduce them to single-component fields, where the spinor structure is encoded in the staggered phase factors. In the following, we will set the lattice spacing to $a=1$ unless otherwise stated.

Using the discretized action, the full path integral can be written as follows:
\begin{equation}
\label{eq:3}
Z  = \int  \prod_x d\bar{\chi}(x)  \, d\chi(x) \, dU_{\mu}(x) \, e^{-S_x[\bar{\chi},\chi,U_{\mu}]}
\end{equation}
One of the main disadvantages of this form is that even though the full action is quadratic in the fermion fields, thus it would be possible to integrate it out as a single determinant, the integral over the external $U_{\mu}(x)$ fields makes it necessary to calculate the determinant for each $U_{\mu}(x)$ configuration \cite{13}. This could quickly make the numerical integration very inefficient for large lattice sizes and higher dimensions \cite{14,15}. The RBF method aims to overcome this issue by separating the fermionic part from the external $U_{\mu}(x)$ fields in the following form:
\begin{equation}
\label{eq:4}
Z \approx \Bigg[  \int \mathcal{D} \bar{\chi} \, \mathcal{D}\chi \, e^{-\tilde{S}[\bar{\chi},\chi]}\Bigg] \times \Bigg[  \int \mathcal{D} U_{\mu} \, F[U_{\mu}]\Bigg] ,
\end{equation}
where $F[U_{\mu}]$ is now a function only of the $U_{\mu}$ fields, while $\tilde{S}[\bar{\chi},\chi]$ depends only on the fermion fields. Technically, this expansion will only serve as a numerical method to approximate the corresponding path integral, e.g., at a fixed gauge, however, if one wants to build a better effective theory, it could also be possible by imposing specific transformation rules on the model parameters.

The way we can approximately achieve the form in Eq.~(\ref{eq:4}) is to apply a radial basis function expansion on the discretized path integral at each lattice site, by expanding $e^{-S}\approx 1-S+\frac{1}{2}S^2+...$ and using the properties of Grassmann variables to obtain a finite number of equations to solve. The training of such a network could be done many ways, however, as the dimensions of the inputs go up, the problem could scale in a nondesired way. In this work, we also propose a numerically affordable method to solve the equations and fit the network parameters even for high dimensions. 

In each example, we will assume periodic boundary conditions to the $U(x)$ fields, and open boundary conditions to the Grassmann variables, so that we do not have additional numerical problems due to the staggered signs and periodic/antiperiodic boundary conditions at even/odd lattice sizes.

\subsection{Staggered fermions in 1D with couplings to on-site external fields}
\label{sec:12}
To show the working principles of the method, let us have the following simplified path integral on a 1D lattice:
\begin{equation}
\label{eq:5}
Z = \int \prod_i d\bar{\chi}_i \, d\chi_i \, dU_i \; e^{-S_i[\bar{\chi},\chi,U]} \, F(U_i),
\end{equation}
where the action $S_i[\bar{\chi},\chi,U]$ is given as
\begin{eqnarray}
\label{eq:7}
S_i[\bar{\chi},\chi,U] = m\bar{\chi}_i \chi_i + \frac{\eta_i}{2}\left( \bar{\chi}_i U_i \chi_{i+1} - \bar{\chi}_i U_i \chi_{i-1} \right)  ,
\end{eqnarray}
where $U_i$ is an external (non-Grassmann) field defined on a finite interval $U_i \in [U_{min},U_{max}]$, while $F(U_i)$ is some function of the $U_i$ fields. From now on we use the notation $\chi_i,U_i$ for the values of the discrete fields at lattice site $i$. This path integral aims to model the staggered fermion action in one dimension where the hopping terms contain couplings to an external $U_i$ field. 

Such discretized action, where the external field only couples to the fermions through $U_i$ and not its neighbors (e.g. $U_{i+1}$), seems to be too simple, however it is not entirely unmotivated see e.g. \cite{4F1,4F2}. In those cases a somewhat similar action can be used to model strongly interacting nonrelativistic fermions, where the four fermion contact interaction term is approximated by the interaction with an external auxiliary scalar field i.e. in the interaction between the Grassmann variables and the scalar fields, we only have $U_i$ on-site scalar terms. It will be shown that by having this assumption, the RBF expansion will be very straightforward, and could be readily applied to such problems without any further complications. 

To proceed further, first we note that this model exhibits the same problem as the full formulation in Eq.\ref{eq:2}, namely that to integrate out for $dU_i$ we need to calculate the fermionic determinant for every $U_i$ field configuration. To overcome this issue, let us apply an RBF expansion for the exponentialized action at each lattice site as follows:
\begin{equation}
\label{eq:Si}
e^{-S_i[\bar{\chi},\chi,U]}\, F(U_i) \approx \sum_{k=1}^K a_k \, e^{-S^{RBF}_{k,i}[\bar{\chi},\chi]} \, e^{-A(U_i-c_k)^2},
\end{equation}
where the pure fermionic part of the action $S^{RBF}_{k,i}[\bar{\chi},\chi]$ is given by
\begin{equation}
\label{eq:S_RBF}
S^{RBF}_{k,i}[\bar{\chi},\chi] = m\bar{\chi}_i \chi_i + \eta_i \frac{b_k}{2}\left( \bar{\chi}_i\chi_{i+1} - \bar{\chi}_i \chi_{i-1} \right)  ,
\end{equation}
where $b_k$, $c_k$, $A$, and $a_k$ are the parameters of the RBF network. The RBF action for the fermionic part has the same Grassmann structure as the original action, which will be important later on. The nonfermionic part $e^{-A(U_i-c_k)^2} $ aims to approximate the residual $U$ dependence from $S_i[\bar{\chi},\chi,U]$ and the $U$ dependence of $F(U_i)$ as well. If the $\chi$ fields were ordinary fields, it would be a very simple task to train the corresponding network, however, as they are Grassmann variables, it is not straightforward on how to train the RBF network. A possible choice is to expand the exponentials in both cases and match the obtained terms order-by-order, which will give us a set of equations that have to be satisfied simultaneously. To see how this works, let us expand the fermionic parts of $e^{-S_i[\bar{\chi},\chi,U]}F(U_i)$ and $\sum_k a_k e^{-S^{RBF}_{k,i}[\bar{\chi},\chi]}e^{-A(U_i-c_k)^2}$ and use the $\chi_i \chi_i=0$ (nilpotency) property of the Grassmann variables \cite{21} to get a finite series, in which case the expansion will stop at linear order and $e^{-S{i}[\bar{\chi},\chi,U]} F(U_i)$ can be given as
\begin{eqnarray}
\label{eq:10}
e^{-S{i}[\bar{\chi},\chi,U]} F(U_i) = \Big(1 - S_i[\bar{\chi},\chi,U]\Big)  F(U_i)
\end{eqnarray}
with $S_i[\bar{\chi},\chi,U] $ given by Eq.~(\ref{eq:Si}). Applying the same idea to the RBF expansion we get:
\begin{eqnarray}
\label{eq:11}
\sum_{k=1}^K a_k \, e^{-S^{RBF}_{k,i}[\bar{\chi},\chi]} \, e^{-A(U_i-c_k)^2} =&&  \\
\sum_{k=1}^K a_k \, \Big(1 - S^{RBF}_{k,i}[\bar{\chi},\chi] \Big)\,  &e^{-A(U_i-c_k)^2}& \nonumber 
\end{eqnarray}
Comparing the different terms in Eq.~(\ref{eq:10}) and Eq.~(\ref{eq:11}), we arrive at two equations that have to be satisfied in order to get a good approximation between the original system and the RBF model:
\begin{equation}
\label{eq:12}
F(U_i) = \sum_{k=1}^K a_k \, e^{-A(U_i-c_k)^2},
\end{equation}
\begin{equation}
\label{eq:13}
U_i \, F(U_i) = \sum_{k=1}^K a_k \, b_k \, e^{-A(U_i-c_k)^2}.
\end{equation}
Solving these two equations is equivalent to training two separate RBF networks, one with linear weights $a_k$ and one with weights $(a_k b_k)$. It can be seen that all the complications from Grassmann variables have vanished, therefore, we can train the networks with ordinary numbers. Before we explain the training procedure, let us go a step further and see what the achievable form is with this kind of expansion. 

The full path integral using the RBF form can be written as a product of the expanded terms at each lattice site as
\begin{eqnarray}
\label{eq:14}
&&Z = \int \prod_i d\bar{\chi}_i d\chi_i dU_i \Bigg[ \sum_{k=1}^K a_k \, e^{- S^{RBF}_{k,i}[\bar{\chi},\chi]}\, e^{-A(U_i-c_k)^2} \Bigg] = \nonumber \\
 &&\sum_{\langle k \rangle \in K^N} \prod_i a_{\langle k\rangle_i}  \Bigg( \left[ \int d\bar{\chi}_i \, d\chi_i \,  e^{-S_{\langle k \rangle_i}[\bar{\chi},\chi]}\right] \times \nonumber \\
&&  \quad \qquad \qquad \qquad \qquad \qquad \left[ \int dU_i \, e^{-A (U_i-c_{\langle k \rangle_i })^2} \right] \Bigg),
\end{eqnarray}
where $\langle k \rangle$ represents all the possible $N$-long combinations of the different coefficients in the RBF expansion, which is coming from the fact that the product of $K$-long sums can be expressed as a large $K^N$ sum, where each term represents one possible combination of the parameter vectors. To make this a bit more clear let us take a very simple example with $K=2$ and $N=3$, in which case one combination of the $a_k$ weights could be $a_{\langle k \rangle} = (a_{\langle k \rangle_1},a_{\langle k \rangle_2},a_{\langle k \rangle_3})=(a_1,a_1,a_2)$. 

The expansion in Eq.~(\ref{eq:14}) in itself starts to look like what we want to achieve, however, due to the very large number of terms for large lattice size, it is insufficient to apply it in this form. The main realization comes from the fact that $U_i$-dependence is defined through an integral in a compact range between $U_{min}$ and $U_{max}$, thus, by using the shifting property of Gaussian integrals, i.e., $\int_a^b e^{-A(x-c)^2} dx \approx \int_{-\infty}^{\infty} e^{-Ax^2}dx$, and by carefully choosing the $A$ widths and $c_k$ centers of the Gaussians, we could get rid of the $k$-dependence in the last line in Eq.~(\ref{eq:14}). This will make it possible to achieve the desired form and make the calculations very efficient. 

To make this more specific let us set:
\begin{equation}
\label{eq:F_U}
F(U_i)=U_i+U_i \sin\left(10\,e^{-20(U_i-0.5)^2} \right)  
\end{equation}
in the range of $U_i \in [0,1]$. This function has multiple local minima in the specified range, and even though it is continuous and bounded from below, a Monte Carlo integration of the corresponding path integral would need very large sample sizes to get good accuracy at larger lattices. 
According to Eq.~(\ref{eq:12}), we have to fit an RBF network that approximates $F(U_i)$ in this range, but we also want to make sure that the following approximation for the Gaussian integrals holds:
\begin{eqnarray}
\label{eq:15}
\int_{0}^{1} dU_i  \, e^{-A(U_i-c_i)^2}  \approx 
\int_{-\infty}^{\infty} dU_i  \, e^{-A U_i^2} ,
\end{eqnarray}
where this equation has to hold for all of the $c_i$ Gaussian centers.
This equation can be approximately satisfied with very good accuracy if we fit the $F(U_i)$ function with a sum of very narrow Gaussian basis functions that are placed inside the predefined interval, which is in this case $[0,1]$. 

There are two criteria that have to be satisfied during optimization. First, the widths and centers have to be set so that at the edges the sum of the Gaussians dies out very fast, which makes sure that we do not introduce extra contributions from outside the physically relevant region. The other criterion is that we have to make sure that between the centers the model gives good generalization and we do not have large oscillations between the training points.

To address these issues, a specific training method has been 'developed' that is able to satisfy the necessary criteria and is very fast even for higher-dimensional problems. The main idea is that we put down $K$ number of Gaussians with a predefined $\Delta_c$ distance from each other that is determined by twice the distance of the half-width point from the centers of the Gaussians and can be determined by $\Delta_c = 2 \sqrt{\ln(2)/A}$. This placement will make sure that intermediate values between the centers will not have large oscillations, thus making it possible to achieve a relatively good generalization with the drawback that it might not correspond to the best possible fit. We will see later that even if the approximation is not perfect, this method is general enough to get good results with below-percentage relative errors. 

In the next step we place down $K$ Gaussians between $[\Delta_c/2, 1-\Delta_c/2]$ with $\Delta_c$ separation from each other, i.e., $[\Delta_c/2, \Delta_c/2+\Delta_c, \Delta_c/2+2\Delta_c, ..., 1-\Delta_c/2]$. To make sure that we have an integer number of Gaussians with this exact distribution, the $A$ width can be parametrized as
\begin{equation}
A = 4 \, k_A^2 \ln(2 )
\end{equation}
where $k_A$ is a positive integer. As $k_A$ gets larger, the Gaussians will be narrower, which means the $\Delta_c/2$ half-width will be smaller too, thus, we get closer and closer to the edges.

After we have determined the positions of the centers, the $a_k$ weight parameters are set considering the value of the approximable function at the predetermined center points $c_k$ as 
\begin{equation}
a_k = C\cdot F(c_k), 
\end{equation}
where $C$ is a fine-tune parameter around $C \approx 1$ that corresponds to the optimal fit for the approximable function in the mean squared sense. This will make sure the small overlaps far from the half-width do not give a large bias to the approximation. For larger dimensions, $C$ tends to be smaller due to the larger overlaps through the more neighboring Gaussians at many dimensions. This value can be calculated using a number of test samples and optimizing $C$ to give the best approximation in the mean squared sense as
\begin{equation}
\label{eq:C_det1}
\arg \min_{\!\!\!\!\!\!\!\!\! \{ C \} } \Bigg\{ \frac{1}{N} \sum_{i=1}^{N} \left( F(\mathbf{U}_i)-C\sum_{k=1}^K a_k e^{-A||\mathbf{U}_i - \mathbf{c}_k ||^2} \right)^2 \Bigg\}  ,
\end{equation}
where $\mathbf{U}_i=(U_i^{(1)},U_i^{(2)}, ... , U_i^{(D)})$ is a vector of the inputs in $D$ dimensions, while $\mathbf{c}_k=(c_k^{(1)},c_k^{(2)}, ... , c_k^{(D)} )$ are the $D$-dimensional centers of the Gaussians.
According to the given description, the fitting procedure can be described in the following three steps:
\\

(1) Choose a large enough $k_A$ value that determines the grid through the number of centers in $D$ dimensions as $n_c=k_A^D$, the width parameters $A= 4k_A^2\ln(2)$, and the distance between the centers as $\Delta_c=2\sqrt{\ln(2)/A}$;
\\

(2) Determine the $C$ fine-tune parameters by optimizing the RBF network to a number of test data using the $F(U)$ function on the full grid;
\\

(3) From the obtained $C$ fine-tune parameters and $c_k$ centers, determine the $a_k$ weights as $a_k = C\, F(c_k)$, where $F(c_k)$ is the approximable function at the predetermined Gaussian centers.
\\

As each chosen $k_A$ value will correspond to a specific accuracy measured by the mean squared error, our task is to determine an optimal $k_A$ that is capable of describing not just the $F(U_i)$ function but its integral as well. We have seen that the RBF approximation will necessarily have some overshoot and undershoot due to the overlapping of the neighboring Gaussians, however, by choosing a 'large enough' $k_A$, these 'oscillations' will be averaged out in the integral.  In general this would mean that $k_A$ should be large enough so that multiple Gaussians could be positioned between the most oscillating parts in the original $F(U_i)$ function so that it could sample those parts as well.

\begin{figure}[!h]
\centering\includegraphics[width=3.2in]{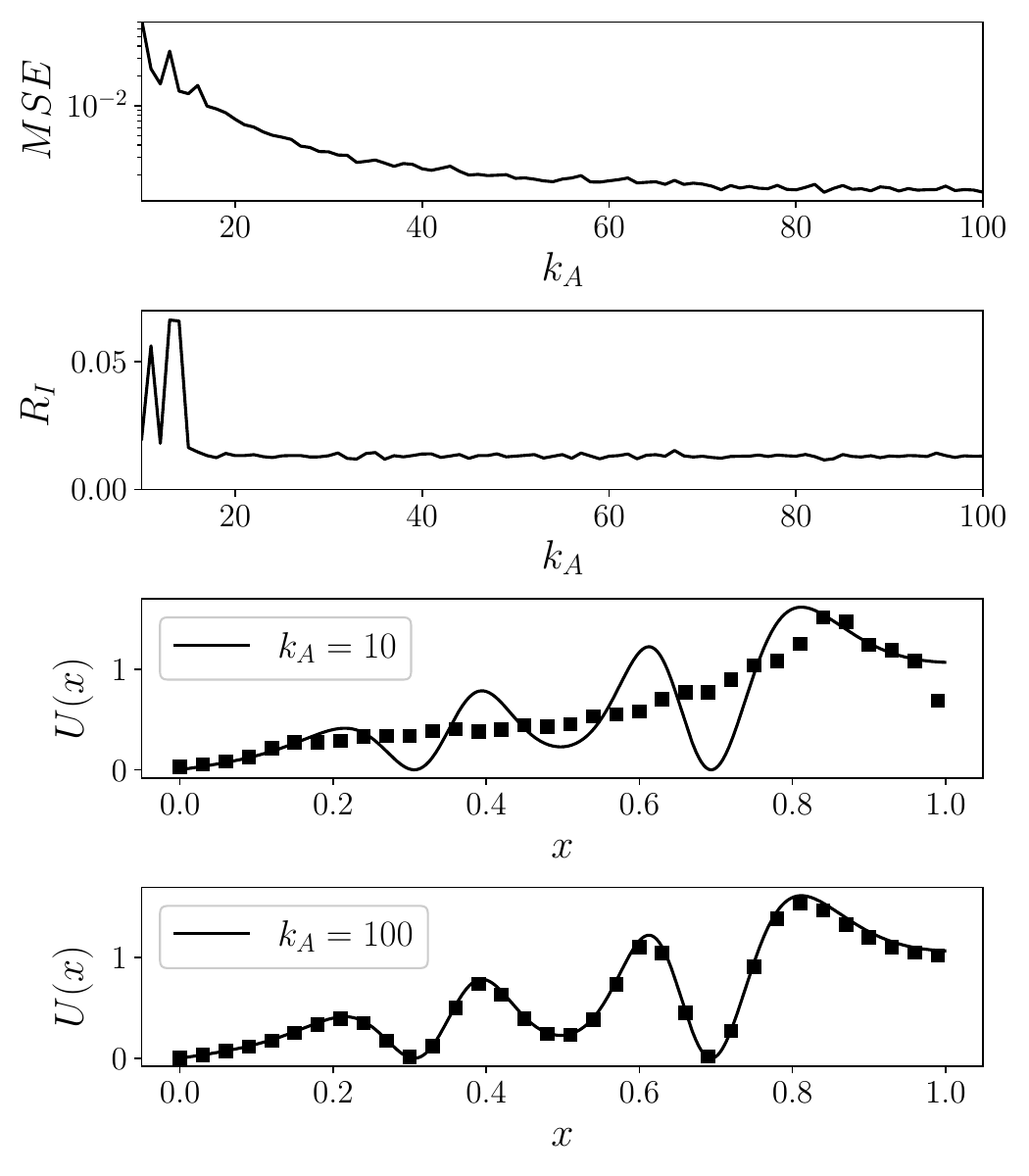}
\caption{Optimization for the $F(U_i) = U_i+U_i \sin\left(10\,e^{-20(U_i-0.5)^2} \right) $ function using the RBF model.  The decreasing $MSE$ suggests that a good pointwise accuracy is achieved for $k_A>40$, while the $R_I$ relative error between the integrals converges to around $1 \%$. The third and fourth plots show the comparison between the true and approximated functions with $k_A=10$ and $k_A=100$, with the optimized $C=0.94$ fine-tune parameter.}
\label{fig:1}
\end{figure}

In Fig.~\ref{fig:1} we show the result of the optimization for the $F(U_i)$ given by Eq.~(\ref{eq:F_U}), using $k_A=10...100$, by comparing the true function to the approximation and their integrals through the relative errors defined as
\begin{equation}
R_I = \frac{\left| \int_{0}^1 dU_i \, F(U_i) - \sum_{k=1}^K a_k\left(\pi/A\right)^{1/2} \right|}{\left| \int_{0}^1 dU_i \, F(U_i) \right|},
\end{equation}
where the number of kernels is given by $K=k_A$, $A=4k_A^2\ln(2)$, $a_k=C \cdot F(c_k)$, with $c_k=[\Delta_c, \Delta_c+2\Delta_c,...,1-\Delta_c]$, and $\Delta_c=2\sqrt{\ln(2)/A}$. We have also used the fact that $\int_{-\infty}^{\infty} dU_i \, e^{-A(U_i-c_k)^2} = (\pi/A)^{1/2}$. Because the $F(U_i)$ function is rather complex and oscillating, the $C$ parameters are determined by using Eq.~(\ref{eq:C_det1}) with the full function on the full grid. 

From the results it can be seen that larger $A$ parameters correspond to an overall better approximation both for the function and for its integral, which is the desired behavior. The relative error of the integral goes down very fast after $k_A>15$, however, to achieve a good pointwise accuracy, it is necessary to use at least $k_A>30$ kernels, which can be seen from the first (upper) plot. After $k_A>30$, the mean squared error still decreases, however, the convergence slows down, which means the RBF approximation was able to sample the most oscillating parts as well. The relative error converges to approximately $1-2 \%$, however, one has to be careful to also monitor the pointwise accuracy through, e.g., the MSE of the function, because the integral in itself is not a good measure for our purposes, and it only serves as a double check. In practice, it is sufficient to choose a $k_A$ value where the decrease of the MSE slows down, i.e., in this case, any $k_A>30-40$ is sufficient. On the bottom two plots we have also shown the approximation for $k_A=10$ and $k_A=100$, showing the previously described behavior. 

After the brief description of the training method that we will use later on, let us proceed further starting from Eq.~(\ref{eq:14}). The $k$-independent approximation of the integral for the $U_i$ fields means that we can set all $c_{\langle k \rangle_i}$ to $0$ and take the integral out from the sum, in which case we arrive at
\begin{equation}
\label{eq:16}
Z \approx I_U^N \cdot \int \prod_i d\bar{\chi}_i d\chi_i \Bigg[ \sum_{k=1}^K a_k \, e^{- S^{RBF}_{k,i}[\bar{\chi},\chi]} \Bigg] ,
\end{equation}
where $I_U^N=\prod_i I_{U_i}$ is the approximated integral using the $U_i$ fields expressed as
\begin{equation}
\label{eq:16_2}
I_{U_i} = \int_{-\infty}^{\infty} dU_i \,e^{-A \, U_i^2} = \left( \frac{\pi}{A}\right)^{\frac{1}{2}}.
\end{equation}
It is important to note that this form was only achievable due to the 'good' parametrization of the RBF network, which means a sufficiently large $A$ parameter (very narrow Gaussians) and a good choice of centers $c_k$ inside the interval $[U_{min},U_{max}]$, which allows us to approximate the $F(U)$ function with a good generalization and small extra contributions from outside the interval. This is now a separated form, however, there is still a product of sum present for the fermionic part. Next, let us expand the exponential in Eq.~(\ref{eq:16}) and gather the terms
\begin{eqnarray}
\label{eq:17}
&&\sum_{k=1}^K a_k \, e^{- S^{RBF}_{k,i}[\bar{\chi},\chi]} = \sum_{k=1}^K a_k \Big( 1 - S_{k,i}^{RBF} \Big) = \\
&&\sum_{k=1}^K a_k \Big( 1 - m\bar{\chi}_i \chi_i - \eta_i\frac{b_k}{2}\left( \bar{\chi}_i\chi_{i+1} - \bar{\chi}_i \chi_{i-1} \right) = \nonumber\\
&&\hat{\mathcal{A}} \Big(  1- m\bar{\chi}_i \chi_i  -\eta_i \frac{\hat{\mathcal{B}}}{2 \hat{\mathcal{A}}}\left( \bar{\chi}_i\chi_{i+1} - \bar{\chi}_i \chi_{i-1} \right) \Big) 
= \hat{\mathcal{A}} \, e^{-\hat{S}_i^{\hat{\mathcal{A}},\hat{\mathcal{B}}}[\bar{\chi},\chi ]} \nonumber 
\end{eqnarray}
where the newly introduced factors $\mathcal{A}$, and $\mathcal{B}$ are given by
\begin{equation}
\label{eq:18}
\hat{\mathcal{A}} = \sum_{k=1}^K a_k , \qquad \qquad \hat{\mathcal{B}}=\sum_{k=1}^K a_k b_k,
\end{equation}
and the new action $S_i^{\hat{\mathcal{A}},\hat{\mathcal{B}}}[\bar{\chi},\chi]$ is expressed as
\begin{equation}
\label{eq:19}
S_i^{\hat{\mathcal{A}},\hat{\mathcal{B}}}[\bar{\chi},\chi] = m \bar{\chi}_i \chi_i + \eta_i \frac{\hat{\mathcal{B}}}{2\hat{\mathcal{A}}}\left( \bar{\chi}_i\chi_{i+1} - \bar{\chi}_i \chi_{i-1} \right).
\end{equation}
This simple form is only allowed by the Grassmann nature of the variables and the fact that the expansion stops at linear order. This form allows us to further simplify Eq.~(\ref{eq:16}) as
\begin{eqnarray}
\label{eq:20}
Z \approx \hat{\mathcal{A}}^N \cdot I_U^N \cdot \int \prod_i d\bar{\chi}_i \, d\chi_i \, e^{-\hat{S}_i^{\hat{\mathcal{A}},\hat{\mathcal{B}}}[\bar{\chi},\chi ]} = \nonumber \\
\hat{\mathcal{A}}^N \, \left( \frac{\pi}{A}\right)^{\frac{N}{2}}   \det(m,\hat{\mathcal{A}},\hat{\mathcal{B}}) ,
\end{eqnarray}
where the fermionic integral is now expressed as the fermionic determinant for the new action. This particularly simple form is able to give a very good approximation to the original path integral written in Eq.~(\ref{eq:5}) for any $F(U_i)$ function that is defined on a finite interval.

The following steps summarize the main method on how to achieve the form in Eq.~(\ref{eq:20}) from Eq.~(\ref{eq:5}):
\\

1. Set up an RBF functional that has the same fermion structure as the original path integral;

2. Expand $e^{-S_i[\bar{\chi},\chi,U]}$, then by matching the coefficients of the Grassmann fields, determine a number of equations that have to be satisfied;

3. Train the RBF networks by choosing 'good' parametrizations for the $A$ widths, and $c_k$ centers;

4. Calculate the $\hat{\mathcal{A}}$, $\hat{\mathcal{B}}$ parameters, then calculate Eq.~(\ref{eq:20}) for a given $N$ (lattice size) and boundary conditions.
\\

To show the capabilities of the model through a numerical example, let us use the same $F(U_i)$ as before, but now with an additional $\alpha$ parameter that changes the shape of the function
\begin{equation}
F_{\alpha}(U_i)=U_i+U_i\sin(\alpha\,e^{-20(U_i-0.5)^2}).
\end{equation}
The corresponding equations of Eq.~(\ref{eq:12}), and Eq.~(\ref{eq:13}) can be satisfied by training $A$, $a_k$, and $b_k$ through the $k_A$, $C_a$, and $C_b$ parameters. According to the previously described training procedure, this means we have to solve the following optimization problem:
\begin{equation}
\arg \!\! \min_{\!\!\!\!\!\!\!\!\! \! !\!\! \! \{k_A,C_a,C_b\}} \Big\{ E_1(k_A,C_a) + E_2(k_A,C_b) \Big\},
\end{equation}
where the first error function $E_1(k_A,C_a)$ corresponds to Eq.~(\ref{eq:12}) and is given by
\begin{equation}
\label{eq:arg1}
E_1(k_A,C_a) \!=\! \frac{1}{N_{T}}\!\sum_{i=1}^{N_{T}} \!\left( \!F_{\alpha}(U_i) - \sum_{k=1}^K \!C_a F_{\alpha}(c_k) \,e^{-A(U_i-c_k)^2 }\! \right)^2 \!\!\!,
\end{equation}
where $N_T$ is the number of training samples, $A=4k_A^2\ln(2)$, and we used the fact that in the model the weight factors are given as $a_k=C_a F_{\alpha}(c_k)$, with $C_a$ serving as a fine-tune parameter that needs to be optimized for a given $A$ width. 

\begin{figure}[!h]
\centering\includegraphics[width=3in]{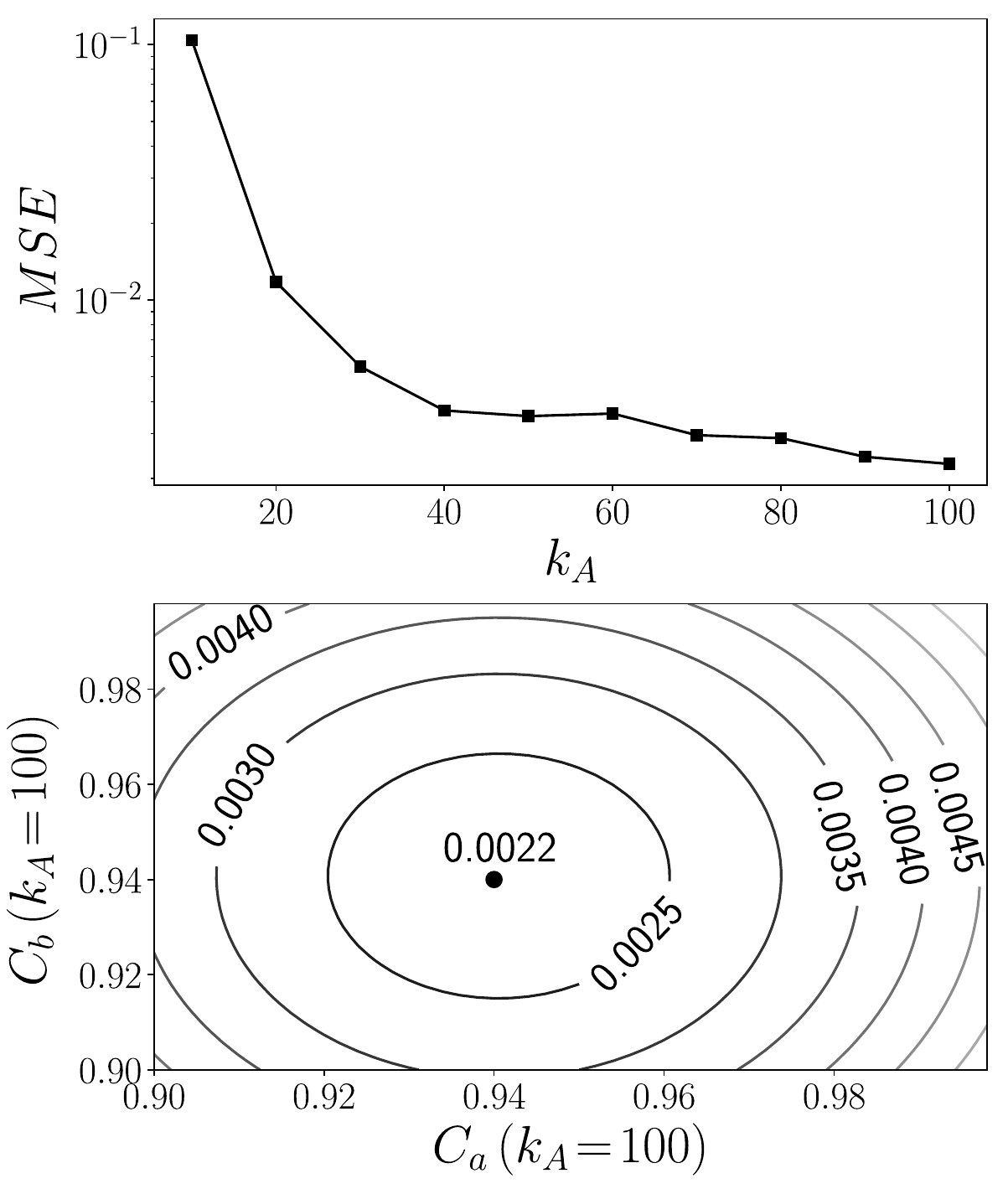}
\caption{Evolution of the MSE with increasing $k_A$ parameter (top), and the corresponding $MSE(C_a,C_b)$ contour plot for $k_A=100$ (bottom). The optimal $C_a$, and $C_b$ fine-tune parameters are $C_a=C_b=0.94$.}
\label{fig:2}
\end{figure}

The second error $E_2(k_A,C_b)$ can be given starting from Eq.~(\ref{eq:13}) by using the fitted $C_a$ and $A$ parameters. The combinations of the weights now should satisfy $a_kb_k=C_b c_kF_{\alpha}(c_k)$, which comes from the constraint that the amplitude of the Gaussians at $c_k$ centers now with $a_k b_k$ amplitudes should be equal to $U_i F_{\alpha}(U_i)$ at $U_i=c_k$ positions.
Here, again, the $C_b$ factor serves as a fine-tune parameter for $a_k b_k$ amplitudes that makes sure that the overlaps at the centers are not giving a large bias. The corresponding mean squared error in this case can be written as
\begin{eqnarray}
\label{eq:arg2a}
E_2(k_A,C_b)= \frac{1}{N_{T}}\sum_{i=1}^{N_{T}} \Bigg( U_i \, F_{\alpha}(U_i) -&& 
 \\
 \sum_{k=1}^K C_b \, c_k \, F_{\alpha}(c_k)&&\, e^{-A(U_i-c_k)^2 }\Bigg)^2 . \nonumber
\end{eqnarray}
In Fig.~\ref{fig:2} we have shown the evolution of the mean squared error using $\alpha=1$, for given $k_A$ values defined as
\begin{equation}
MSE(k_A) = E_1(k_A,C_a^{*}) + E_2(k_A,C_b^{*}),
\end{equation}
where $C_a^{*}$, and $C_b^{*}$ are the optimized fine-tune parameters at a specific $k_A$.
It can be seen that the $MSE(k_A)$ is decreasing with larger $k_A$ (i.e., with increasing number and narrower Gaussians). After $k_A>40$ the decrease slows down, which means any $k_A>40$ parameter is sufficient. Nevertheless, in this example we will choose $k_A=100$, in which case, on the bottom figure, we show the contour plot of the mean squared error for the case of a fine grid of $C_a$ and $C_b$ values. The results show that the optimal parameters are $C_a^{*}=C_b^{*}=0.94$, which value will be used for all the other functions with different $\alpha$ values as well.

After we have set the model, the path integral can be approximated for any specific $N$ through Eq.~(\ref{eq:20}), with $\hat{\mathcal{A}}$ and $\hat{\mathcal{B}}$ parameters given as
\begin{equation}
\hat{\mathcal{A}} = C_a \sum_{k=1}^K  F_{\alpha}(c_k), \qquad \qquad \hat{\mathcal{B}}=C_b \sum_{k=1}^K c_k F_{\alpha}(c_k).
\end{equation}

\begin{figure}[!h]
\centering\includegraphics[width=3in]{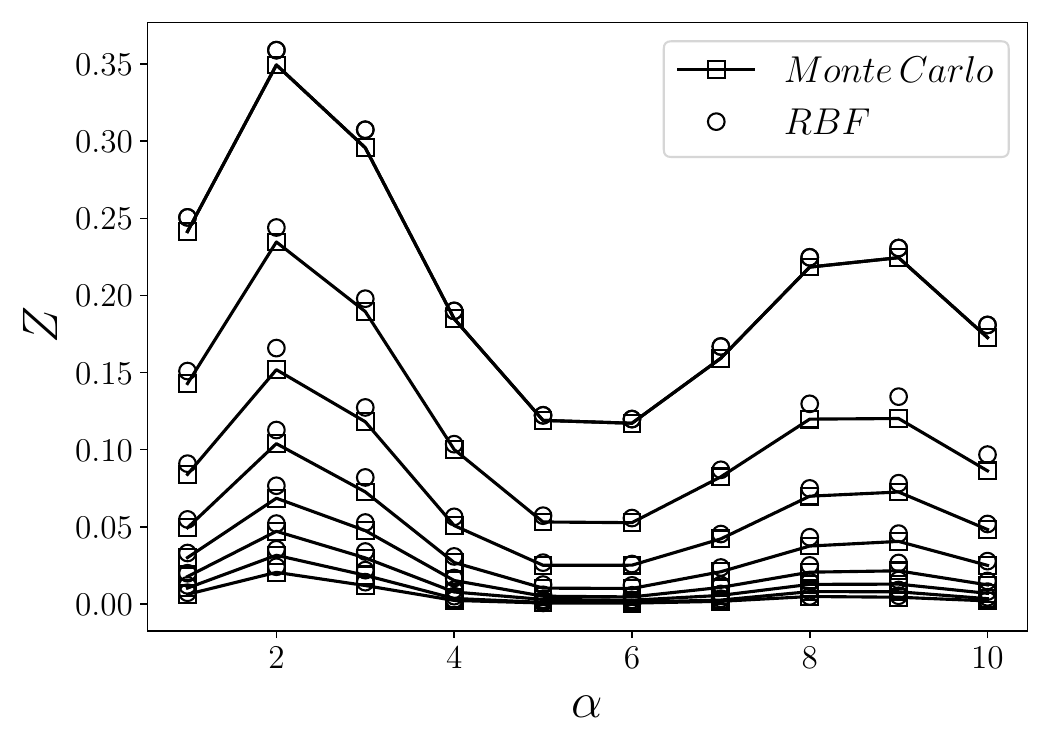}
\caption{Comparison of the results of the path integral using Monte Carlo integration, and the RBF model for different lattice sizes and $\alpha$ parameters, using $F_{\alpha}(U_i)=U_i+U_i\sin(\alpha\,e^{-20(U_i-0.5)^2})$. }
\label{fig:3}
\end{figure}

In Fig.~\ref{fig:3} we show the comparison between the true values of $Z$ (calculated by Monte Carlo integration), and the approximation that is given by the RBF model for $\alpha=[1,2,3,4,5,6,7,8,9,10]$ and $N=[3,4,5,6,7,8,9,10]$. In these calculations the mass is set to $m=1$, and we use open boundary conditions.
The comparison shows that indeed a very good, few percent accuracy can be achieved by using the RBF expansion. The small discrepancies are coming from both the uncertainty of the Monte Carlo estimate, and from the inherent uncertainty of the RBF expansion that coming from e.g. the determination of the $C_a, C_b$ fine tune parameters.

There are a few important notes to be considered. First, the main simplification in the path integral in this example was the fact that the hopping terms only coupled to on-site $U_i$ external fields, which is not necessarily a relevant application if one wants to study, e.g., gauge theories, in which case we would need a coupling to $U_{i-1}$ as well. However, this form was perfectly fine to show the working principles of the model when there is a local coupling to an external field through the derivative terms, which would make the full calculation problematic otherwise. 

The second thing to consider is the specific training method we have applied to solve the relevant equations that appeared through expanding the exponentialized fermionic action. In low-dimensional problems, other gradient-based methods \cite{22,23,24}, where we also train the centers of the Gaussians, could also give a very versatile and fast way to solve the equations, however, as we introduce more input parameters into the model, the optimization will be harder to solve with each new dimension introduced. 

The third comment is regarding the intervals for the $U$ fields, which was set to $[0,1]$ in our example before. Of course $U_{min}$ and $U_{max}$ could be any other value as well, however, by scaling and/or shifting the $U$ fields, it is possible to normalize it between $[0,1]$ by also changing the $F(U)$ function and the measure of the integral. This, of course, needs to be done carefully, especially in the case of gauge fields, where $dU$ corresponds to the Haar measure of the specific group.

In the next subsection we will extend the model with the inclusion of neighboring site $U_{i-1}$ couplings in the hopping terms. This will make things more complicated, however, it will be shown that a good approximation can still be achieved by carefully analyzing the overlapping integrals that will necessarily arise.

\subsection{Staggered fermions in 1D with couplings to on, and neighboring site external fields}
\label{sec:13}
In this section we will generalize the model to the case when the hopping terms are coupled not just to $U_i$, but to $U_{i-1}$ as well. The corresponding action will now have the form of
\begin{eqnarray}
\label{eq:21}
\!\!\!\! \!\!\!\! S_i[\bar{\chi},\chi,U] = m\bar{\chi}_i \chi_i + \frac{\eta_i}{2}\left( \bar{\chi}_i U_i \chi_{i+1} - \bar{\chi}_i U_{i-1} \chi_{i-1} \right)  ,
\end{eqnarray}
with the full path integral,
\begin{equation}
\label{eq:22}
Z = \int \prod_i d\bar{\chi}_i \, d\chi_i \, dU_i \; F(U_i,U_{i-1}) \, e^{-S_i[\bar{\chi},\chi,U]},
\end{equation}
where $F(U_i,U_{i-1})$ is some function of the external fields. Note that this function could have contained other neighbors of $U_i$ as well e.g. $U_{i+1}$, however by using only the fields that also couple to the Grassmann variables does not take away from the generality of the model. The reasoning behind this will be further discussed later on. 
This form is now one step closer to physically relevant models, and the technique that will be shown here can be directly applied in more meaningful theories. 

Starting from the RBF expansion that is described in the previous section, we will expand $F(U_i,U_{i-1})e^{-S_i[\bar{\chi},\chi,U]}$ as a radial basis function network that has the same fermionic structure as the original action, but now the nonfermionic part has to include $U_i$, and $U_{i-1}$ as input parameters as follows:
\begin{eqnarray}
\label{eq:23}
&&F(U_i,U_{i-1})\,e^{-S_i[\bar{\chi},\chi,U]} \approx \sum_{k=1}^K a_k \, e^{-S^{RBF}_{k,i}[\bar{\chi},\chi]} \times \nonumber \\
&& \qquad \qquad \qquad e^{-A\Big[ (U_i-c_{k}^{(1)})^2  + (U_{i-1}-c_{k}^{(2)})^2\Big]}, 
\end{eqnarray}
where the RBF fermionic action is now given as
\begin{equation}
\label{eq:24}
S^{RBF}_{k,i}[\bar{\chi},\chi] = m\bar{\chi}_i \chi_i + \frac{\eta_i}{2}\left( b_{k}^{(1)}\bar{\chi}_i\chi_{i+1} - b_{k}^{(2)}\bar{\chi}_i \chi_{i-1} \right)  ,
\end{equation}
where, due to the different couplings to $\bar{\chi}_i \chi_{i+1}$ and $\bar{\chi}_i \chi_{i-1}$, we need to introduce two RBF parameters $b_{k}^{(1)}$ and $b_{k}^{(2)}$ for each term.  Accordingly, due to the dependence on $U_i$ and $U_{i-1}$ at site $i$, the $U$-dependent term in Eq.~(\ref{eq:23}) needs two inputs, thus, the expansion effectively becomes a two-dimensional RBF network with inputs $U_{i}$, $U_{i-1}$, and parameters $A$, $c_{k}^{(1)}$, $c_{k}^{(2)}$, $b_{k}^{(1)}$, and $b_{k}^{(2)}$. We will see that even though the number of parameters increased, the problem itself will not become harder to solve with the specific training method that is described before.

After expanding the exponential of the fermionic part in the original and in the RBF approximation and matching the coefficients, we arrive at the following three equations:
\begin{equation}
\label{eq:25}
F(U_i,U_{i-1}) = \sum_{k=1}^K a_k\,e^{-A\Big[ (U_i-c_{k}^{(1)})^2  + (U_{i-1}-c_{k}^{(2)})^2\Big]}
\end{equation}
\begin{equation}
\label{eq:26}
F(U_i,U_{i-1})\,U_i = \sum_{k=1}^K a_k\, b_{k}^{(1)}e^{-A\Big[ (U_i-c_{k}^{(1)})^2  + (U_{i-1}-c_{k}^{(2)})^2\Big]}
\end{equation}
\begin{equation}
\label{eq:27}
F(U_i,U_{i-1})\,U_{i-1}= \sum_{k=1}^K a_k b_{k}^{(2)}\, e^{-A\Big[ (U_i-c_{k}^{(1)})^2  + (U_{i-1}-c_{k}^{(2)})^2\Big]},
\end{equation}
where we have the same chain of linear equations that we need to optimize for the $a_k$, $A$, $b_k$, and $c_k$ parameters for two-dimensional functions that are spanned in the $(U_i, U_{i-1})$ space. Like in the previous example, we can start from the first of these equations [Eq.~(\ref{eq:25})] and fit the corresponding $a_k$, $A$, and $c_k$ parameters, then proceed to Eq.~(\ref{eq:26}) and Eq.~(\ref{eq:27}) and fit the remaining $b_k$ parameters separately. Using the specific training method that we have described in the previous section, we only need to fit four parameters ($A$, $C_a$, $C_{b}^{(1)}$, $C_b^{(2)}$), where the fine-tuned parameters are defined through,
\begin{eqnarray}
\label{eq:28}
a_k&=&C_a \, F(c_k^{(1)},c_k^{(2)}),\nonumber \\
a_k b_k^{(1)}&=&C_b^{(1)} c_k^{(1)} \, F(c_k^{(1)},c_k^{(2)}), \\
a_k b_k^{(2)}&=&C_b^{(2)} c_k^{(2)} \, F(c_k^{(1)},c_k^{(2)}), \nonumber
\end{eqnarray}
where again $c_k^{(1)}$ and $c_k^{(2)}$ centers are defined through the $A$ width parameter of the basis functions that defines the $\Delta_c$ distances through the half-width of the Gaussians.  The corresponding optimization problem thus can be formulated as follows:
\begin{equation}
\label{eq:ex2_1}
\arg \!\!\!\!  \min_{\!\!\!\!\!\!\!\!\! \left\{ \substack{k_A,C_a \\ C_b^{(1)}\!\!,C_b^{(2)}} \right\}}
\Big\{ E_1(k_A,C_a) + E_2(k_A,C_b^{(1)}) + E_3(k_A,C_b^{(2)}) \Big\},
\end{equation}
where now we have three separate error terms $E_1$, $E_2$, and $E_3$ that have to be determined for each $k_A$, and can be given as

\begin{eqnarray}
\label{eq:29}
&&E_1(k_A,C_a)  = \Bigg\{\frac{1}{N_{T}}\sum_{i=1}^{N_{T}} \Bigg( F(U_i,U_{i-1}) - \\ \sum_{k=1}^K &&\,C_a  F(c_k^{(1)},c_k^{(2)})\,e^{-A\Big[ (U_i-c_{k}^{(1)})^2  + (U_{i-1}-c_{k}^{(2)})^2\Big]}\Bigg)^2 \Bigg\}, \nonumber
\end{eqnarray}
\begin{eqnarray}
\label{eq:30}
&&E_2(k_A,C_b^{(1)}) =  \Bigg\{\frac{1}{N_{T}}\sum_{i=1}^{N_{T}} \Bigg( F(U_i,U_{i-1})\,U_i \,- \\ 
&&\sum_{k=1}^K \,C_b^{(1)}c_k^{(1)} F(c_k^{(1)},c_k^{(2)}) \,e^{-A\Big[ (U_i-c_{k}^{(1)})^2  + (U_{i-1}-c_{k}^{(2)})^2\Big]}\Bigg)^2 \Bigg\}, \nonumber
\end{eqnarray}
\begin{eqnarray}
\label{eq:31}
&&E_3(k_A,C_b^{(2)})  = \Bigg\{\frac{1}{N_{T}}\sum_{i=1}^{N_{T}} \Bigg( F(U_i,U_{i-1})\,U_{i-1} \,- \\ 
&&\sum_{k=1}^K \,C_b^{(2)}c_k^{(2)}  F(c_k^{(1)},c_k^{(2)})\,e^{-A\Big[ (U_i-c_{k}^{(1)})^2  + (U_{i-1}-c_{k}^{(2)})^2\Big]}\Bigg)^2 \Bigg\}, \nonumber
\end{eqnarray}
After solving the optimization problem for $k_A$, $C_a$, $C_b^{(1)}$, and $C_b^{(2)}$, we can obtain the necessary $a_k$ and $b_k$ parameters using Eq.~(\ref{eq:28}). 

There is one arising complication, however, due to the inclusion of $U_{i-1}$, which will be apparent by writing out the product of sums according to Eq.~(\ref{eq:14}) as follows:
\begin{eqnarray}
\label{eq:32}
&&Z \approx \!\!\! \sum_{\langle k \rangle \in K^N} \prod_i a_{\langle k \rangle_i}  \Bigg( \left[ \int d\bar{\chi}_i \, d\chi_i \,  e^{-S_{\langle k \rangle_i}[\bar{\chi},\chi]}\right] \times \nonumber \\
&& \quad \qquad \ \left[ \int dU_i \, e^{-A\Big[ (U_i-c_{\langle k \rangle_i}^{(1)})^2  + (U_{i-1}-c_{\langle k \rangle_i}^{(2)})^2\Big]} \right] \Bigg),
\end{eqnarray}
where again the sum corresponds to all the possible combinations of the parameter vectors. The reason we were able to simplify this form to Eq.~(\ref{eq:20}) is because the $dU_i$ integral on the second line could be approximated by a $k$-independent form, thus, we could take the full $U$-dependent part outside the sum. In this case, due to the $U_{i-1}$ terms, this factorization is not that straightforward. 

After reorganizing the terms, the integrals defined this way can be cast in the form of $\left(\frac{\pi}{2A}\right)^{1/2} e^{-\delta_i}$, where the $\delta_i$ suppression factor depends on the width of the Gaussians and the distances between the centers, thus, only a fraction of the terms will dominate the full integral, e.g., the dominant term is when all $\delta_i=0$. 
Instead of counting the dominant terms in the full integral, we will use a different route, and use the statistical properties of the large sums that we have obtained after taking the full product over all lattice sites. 

First, assuming periodic boundary conditions for the $U(x)$ fields, we can reorganize Eq.~(\ref{eq:32}) so that the integral of $U(x)$ at each site is separated as follows:
\begin{equation}
\label{eq:33}
Z \approx \sum_{\langle  k \rangle \in K^N } d_{\langle k \rangle}\, \prod_i \, \int dU_i \, e^{-A\Big[ (U_i-c_{\langle k \rangle_i}^{(1)})^2  + (U_{i}-c_{\langle k \rangle_{i+1}}^{(2)})^2\Big]},
\end{equation}
where $d_{\langle k \rangle}$ represents a fermionic determinant that comes from the product in Eq.~(\ref{eq:32}) and is given by
\begin{equation}
\label{eq:34}
d_{\langle k \rangle} = \prod_i a_{\langle k \rangle_i}  \Bigg( \left[ \int d\bar{\chi}_i \, d\chi_i \,  e^{-S_{\langle k \rangle_i}[\bar{\chi},\chi]}\right].
\end{equation}
Note that the product of the Gaussians now have the form of $\prod_i G(U_i,c^{(1)}_{\langle k \rangle_i},c^{(2)}_{\langle k \rangle_i})$ with $U_i$ and both centers placed between $0$ and $1$. When $A$ is large enough this will allows us to approximate the integral as
\begin{equation}
\label{eq:35}
Z \approx \sum_{\langle  k \rangle \in K^N } d_{\langle k \rangle}\, \prod_i \, 
\left( \frac{\pi}{2A}\right)^{\frac{1}{2}} e^{-\frac{A}{2} ( c^{(1)}_{\langle k \rangle_i} - c^{(2)}_{\langle k \rangle_{i+1}})^2},
\end{equation}
where we have neglected the small contributions at the boundaries near $0$ and $1$. In Appendix A it is shown in detail, that even though $d_{\langle k \rangle}$ depends on the centers through $a_{\langle k \rangle_i}$ and the fermion action, Eq.~(\ref{eq:35}) can be approximated by the following factorized form:
\begin{eqnarray}
\label{eq:36}
&&Z \approx \frac{1}{K^N} \left( \frac{\pi}{2A} \right)^{\frac{N}{2}} \left[ \sum_{\langle  k \rangle \in K^N } d_{\langle k \rangle}\right] \times \nonumber \\
&& \qquad \left[ \sum_{\langle  k \rangle \in K^N } \prod_i \, e^{-\frac{A}{2} ( c^{(1)}_{\langle k \rangle_i} - c^{(2)}_{\langle k \rangle_{i+1}})^2}\right] \times R^N,
\end{eqnarray}
where $R$ is asymptotically a constant coefficient that depends on the actual form of $d_{\langle k \rangle}$ i.e. on the fermion determinant, and on the $F(U_i,U_{i-1})$ function. This factorization is possible due to the special form of the RBF approximation, and is related to the fact that the covariance between the Gaussians and the $d_{\langle k \rangle}$ is small compared to their expectation values. In practice $R\approx 1$, but can be smaller or larger depending on the actual functional form of $d_{\langle k \rangle}$. In Appendix A, we further investigate this factorization, and show that it is quite general and is applicable to other type of systems as well e.g. Dirac fermions.

Assuming that the factorization works, we can proceed further by realizing that due to the equidistant placement of the $c_k$ centers, the distribution of $c_{\langle k \rangle_i}$ for large $K$ and $N$ will be uniform on the same interval as $c_k$. This will make it possible to write the large sum over $c_{\langle k \rangle_i}$ as an integral over $dc_i$ for each lattice site. Using this to the sum of Gaussians in Eq.~(\ref{eq:36}) the path integal can be written as
\begin{eqnarray}
\label{eq:37}
&&Z \approx \frac{R^N}{K^N}\left( \frac{\pi}{2A}\right)^{\frac{N}{2}} \left[ \sum_{\langle  k \rangle \in K^N } d_{\langle k \rangle}\right] \times \nonumber \\
&&\qquad K^N \left[\prod_i \int d c_{\langle k \rangle_i }^{(1) } d c_{\langle k \rangle_i }^{(2) } \, 
e^{-\frac{A}{2} ( c^{(1)}_{\langle k \rangle_i} - c^{(2)}_{\langle k \rangle_{i+1}})^2}\right]  \approx \nonumber \\
&&\qquad R^N\left( \frac{\pi}{A}\right)^{N} \left[ \sum_{\langle  k \rangle \in K^N } d_{\langle k \rangle}\right]
\end{eqnarray}
where we have used the fact that the centers are distributed uniformly between $0$ and $1$, and $\int dc_1 dc_2 \, e^{-\frac{A}{2}(c_1-c_2)^2} \approx (2\pi/A)^{1/2}$ for sufficiently large $A$. In the case when the range of the centers are different a corresponding scaling has to be applied.

The factorized form in Eq.~(\ref{eq:37}) allows us to rewrite the $\sum_{\langle k \rangle } d_{\langle k \rangle}$ sum back into a product of $K$-sums, in which case we arrive at the following form:
\begin{equation}
\label{eq:39}
Z \approx R^N \left( \frac{\pi}{A}\right)^N  \int \prod_i d\bar{\chi}_i d\chi_i \Bigg[ \sum_{k=1}^K a_k e^{-S_{k,i}^{RBF}[\bar{\chi},\chi]}\Bigg],
\end{equation}
where $S_{k,i}^{RBF}[\bar{\chi},\chi]$ is given by Eq.~(\ref{eq:24}).
The fermionic part is very similar to the one shown in Eq.~(\ref{eq:17}) with the distinction that it now has $C_a$, $C_b^{(1)}$, $C_b^{(2)}$. $b_{k}^{(1)}$, $b_{k}^{(2)}$ parameters. According to the same idea, the fermionic sum can be recast into the following simple form:
\begin{equation}
\label{eq:41}
\sum_{k=1}^K a_k e^{-S_{k,i}^{RBF}[\bar{\chi},\chi]} = \hat{\mathcal{A}}\, e^{-\hat{S}_i^{\hat{\mathcal{A}},\hat{\mathcal{B}}_1,\hat{\mathcal{B}}_2}[\bar{\chi},\chi]},
\end{equation}
where
\begin{equation}
\label{eq:42}
S_i^{\hat{\mathcal{A}},\hat{\mathcal{B}}}[\bar{\chi},\chi] = m \bar{\chi}_i \chi_i + \frac{\eta_i}{2\hat{\mathcal{A}}}\left( \hat{\mathcal{B}}_1\bar{\chi}_i\chi_{i+1} - \hat{\mathcal{B}}_2\bar{\chi}_i \chi_{i-1} \right),
\end{equation}
with the coefficients set to
\begin{eqnarray}
\label{eq:43}
&&\hat{\mathcal{A}} = C_a \sum_{k=1}^K  F(c_k^{(1)},c_k^{(2)}), \\
&&\hat{\mathcal{B}}_1=C_b^{(1)} \sum_{k=1}^K c_k^{(1)} F(c_k^{(1)},c_k^{(2)}), \\
&& \hat{\mathcal{B}}_2=C_b^{(2)} \sum_{k=1}^K c_k^{(2)} F(c_k^{(1)},c_k^{(2)}).
\end{eqnarray}

By putting everything together, the RBF approximation of the path integral can be cast into the following form:
\begin{eqnarray}
\label{eq:44}
&& \!\!\!\!\!  \!\!\!\!\!  \!\!\!\!\! Z \approx  R^N \left( \frac{\pi}{A}\right)^{N}  \hat{\mathcal{A}} ^{N} \prod_i d\bar{\chi}_i d\chi_i e^{-\hat{S}_i^{\hat{\mathcal{A}},\hat{\mathcal{B}}_1,\hat{\mathcal{B}}_2}} =  \nonumber \\
&& \qquad = R^N \left( \frac{\pi}{A} \right)^{N}  \hat{\mathcal{A}}^{N} \det(m,\hat{\mathcal{A}},\hat{\mathcal{B}}_1, \hat{\mathcal{B}}_2).
\end{eqnarray} 
In practice the absolute value of $Z$ is not necessarily an important quantity, and we are mostly interested in its logarithm $\ln Z$ and its evolution and ratios, rather than its actual values. 
The $R$ parameter could be approximated by a direct evaluation of the factorization formula in Eq.~(\ref{eq:36}), however it is more straightforward to fit it through evaluating the actual path integral at small lattice sizes, in which case the possible small boundary effects that were neglected in the derivation can also be accounted for.

To test the method, let us set the mass $m=1$, the intervals $U_{min}=0$ and $U_{max}=1$, and the $F$ function to:
\begin{equation}
F(U_i,U_{i-1}) = e^{U_i U_{i-1}},
\end{equation}
where again periodic boundary condition is assumed for $U_i$, and open boundary condition for the $\chi_i$, $\overline{\chi}_i$ Grassmann variables in the determinant.

\begin{figure}[!h]
\centering\includegraphics[width=3in]{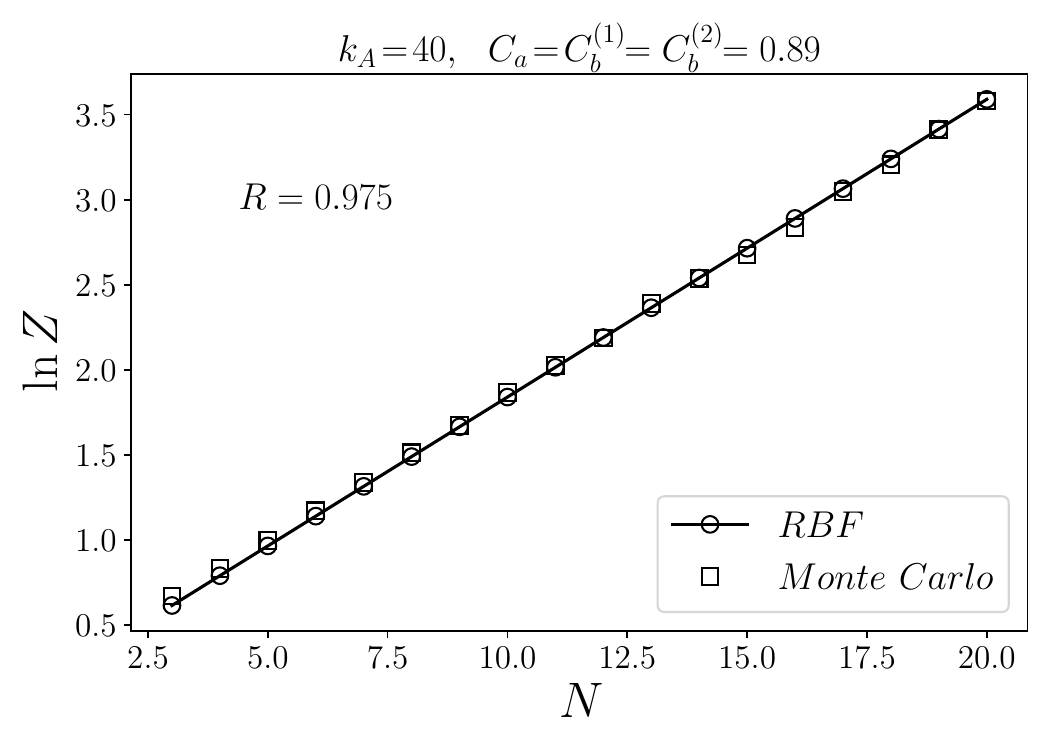}
\caption{Comparison of the Monte Carlo results, with the RBF approximation at different lattice sizes between $N\!=\!3$ and $N\!=\!20$ for the system given by Eq.~(\ref{eq:22}), with $F(U_i,U_{i-1})=e^{U_iU_{i-1}}$, and $m=1$. The factorization parameter has been fitted through the Monte Carlo points at $N=3$ and $N=4$ to $R \approx 0.975$, using the RBF parameters $k_A=40$, $C_a=0.89$, and $C_b^{(1)}=C_b^{(2)}=0.89$.
}
\label{fig:21}
\end{figure}

Using this functional form, we have solved the corresponding optimization problem shown in Eq.~(\ref{eq:28}), Eq.~(\ref{eq:29}), and (\ref{eq:30}) with $k_A=40$, to obtain $C_a=0.89$, $C_b^{(1)}=0.89$, and $C_b^{(2)}=0.89$. The $R$ parameter is fitted using the value of $Z_{true}$ at $N=3$, and $N=4$, giving $R\approx 0.975$. The comparison of $\ln Z$ between the RBF approximation and the true values obtained by Monte Carlo integration can be followed in Fig.~\ref{fig:21} up to $N=20$, showing again few percent accuracy between the true values and the approximations. 

\begin{figure}[!h]
\centering\includegraphics[width=3in]{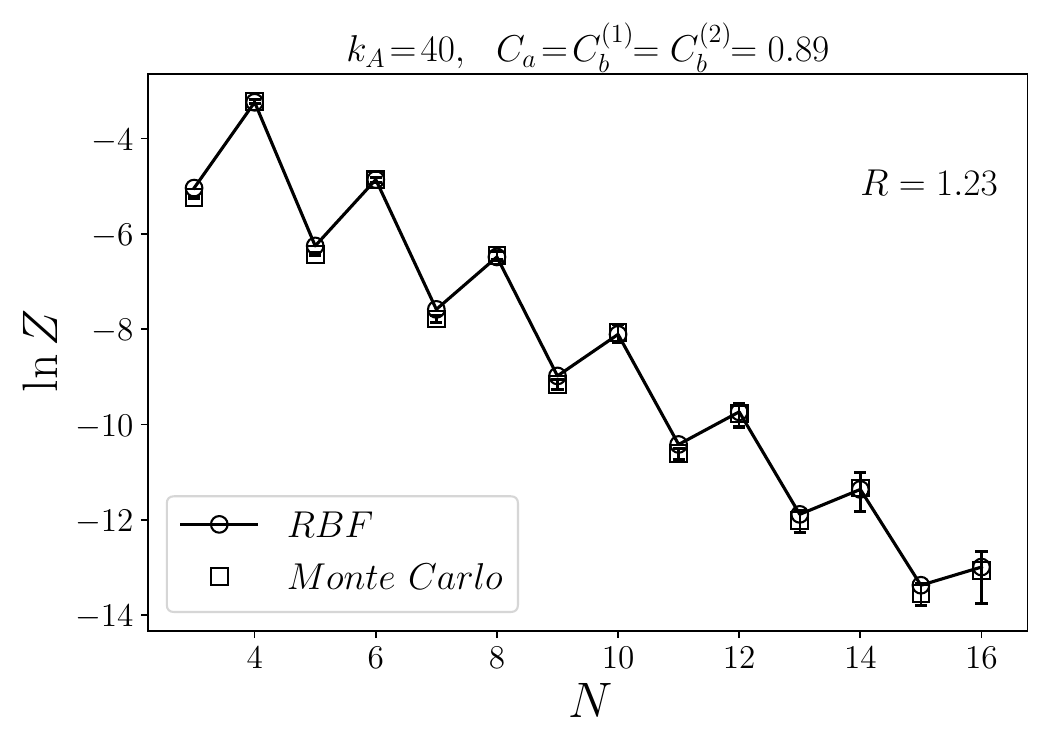}
\caption{Comparison of the Monte Carlo results, with the RBF approximation at different lattice sizes between $N\!=\!3$ and $N\!=\!16$ for the system given by Eq.~(\ref{eq:22}), with $F(U_i,U_{i-1})=e^{U_iU_{i-1}}$, and $m=0.01$. The factorization parameter has been fitted through the Monte Carlo points at $N=3$ and $N=4$ to $R \approx 1.23$, using the RBF parameters $k_A=40$, $C_a=0.89$, and $C_b^{(1)}=C_b^{(2)}=0.89$.
}
\label{fig:22}
\end{figure}

Before we show that the model can be easily extended to higher dimensions, let us do another calculation for the same system, but now with a smaller mass set to $m=0.01$. In this case the $R$ parameter is fitted to $R \approx 1.23$, and the results can be seen in Fig.~\ref{fig:22}.
As it can be seen, the RBF model is again very close to the Monte Carlo results even for larger lattice sizes.

\section{Extensions to higher dimensions}
\label{sec:2}
In the previous section we have shown that the RBF expansion is able to give a very good approximation to the path integral in one dimension. In this section, we will show that the exact same method can be readily applied to higher space-time dimensions and for more complicated models as well. 

The model will be extended to 1+1 dimensions in which case the staggered action takes the form of
\begin{eqnarray}
\label{eq:3_1}
&&S_{ij}[\bar{\chi},\chi,U^{(\mu)}] = m\bar{\chi}_{ij}\chi_{ij} + \eta_{ij}^{(0)}  \Bigg[ \frac{1}{2} \Big( \bar{\chi}_{ij} U^{(0)}_{ij} \chi_{i+1,j} -
\nonumber \\
&&
\quad \bar{\chi}_{ij} U^{ (0)}_{i-1,j} \chi_{i-1,j}  \Big)  \Bigg] + 
\eta_{ij}^{(1)}  \Bigg[ \frac{1}{2} \Big( \bar{\chi}_{ij} U^{(1)}_{ij} \chi_{i,j+1} -
\nonumber \\
&&
\quad \bar{\chi}_{ij} U^{ (1) }_{i,j-1} \chi_{i,j-1}  \Big)  \Bigg] ,
\end{eqnarray}
where $U^{(\mu)}_{ij}$ represents the external fields (e.g. gauge fields) in the $\mu$ direction, while $\eta^{(\mu)}_{ij}$ are the corresponding staggered fermion sign factors. 
The corresponding path integral now reads as
\begin{equation}
\label{eq:3_3}
Z = \int \prod_{ij} d\bar{\chi}_{ij}\, d\chi_{ij}\, dU^{(0)}_{ij}\, dU^{(1)}_{ij}F(U^{(\mu)})\, e^{-S_{ij}[\bar{\chi},\chi,U^{(\mu)}]},
\end{equation}
where $F(U^{(\mu)})$ can be a function of $U_{ij}^{(\mu)}$, $U_{ij}^{(\mu)\dagger}$ and its neighbors as before. For simplicity, we will assume that $F(U^{(\mu)})$ only depends on the variables that are present at the hopping terms in the fermionic determinant i.e. $U_{ij}^{(0)}$, $U_{i-1,j}^{(0)}$, $U_{ij}^{(1)}$, $U_{i,j-1}^{(1)}$.
This simplification does not take away from the generality of the model, as it will only increase the input dimensions [due to extra neighboring $U^{(\mu)}$ terms] and change the form of the equations that have to be solved. While the equations will be slightly more complicated, the same methodology can be readily applied.

As before, the RBF approximation starts from expanding $F(U^{(\mu)})\,e^{-S_{ij}[\bar{\chi},\chi,U^{(\mu)}]}$ using radial basis functions as kernels. In this case the input dimension of the RBF model will be 4, due to the dependence on $U_{ij}^{(0)}$, $U_{ij}^{(1)}$, $U_{i-1,j}^{(0)}$, and $U_{i,j-1}^{(1)}$, and can be written as
\begin{eqnarray}
\label{eq:3_4}
&&F(U^{(\mu)})\,e^{-S_{ij}[\bar{\chi},\chi,U^{(\mu)}]} \approx \sum_{k=1}^K a_k \, e^{-S_{k,ij}^{RBF}[\bar{\chi},\chi]} \times \\
&&e^{-A \Big[ (U_{ij}^{(0)}-c_k^{(1)})^2 + (U_{ij}^{(1)}-c_k^{(2)})^2 + (U_{i-1,j}^{(0)}-c_k^{(3)})^2 + (U_{i,j-1}^{(1)}-c_k^{(4)})^2   \Big]} \nonumber,
\end{eqnarray}
where the fermionic action is now given by the following form:
\begin{eqnarray}
\label{eq:3_5}
&&S_{k,ij}^{RBF}[\bar{\chi},\chi] = m\bar{\chi}_{ij}\chi_{ij} + \eta_{ij}^{(0)}  \Bigg[ \frac{1}{2} \Big(b_k^{(1)} \bar{\chi}_{ij}\chi_{i+1,j} -
 \\
&&b_k^{(2)} \bar{\chi}_{ij}  \chi_{i-1,j}  \Big)  \Bigg] \!\!+\! 
\eta_{ij}^{(1)}  \Bigg[ \frac{1}{2} \Big( b_k^{(3)} \bar{\chi}_{ij} \chi_{i,j+1} - b_k^{(4)} \bar{\chi}_{ij} \chi_{i,j-1}  \Big)  \Bigg] . \nonumber
\end{eqnarray}
Note that the $b_k^{(1)}$,$b_k^{(2)}$,$b_k^{(3)}$, and $b_k^{(4)}$ parameters can be complex as well, in which case they can be written as
$b_k^{(n)} =  b_{A,k}^{(n)}\, e^{2\pi i b_{\phi,k}^{(n)}}$. This representation would allows us to include adjugate variables into the hopping terms, that would be necessary in describing e.g. U(1) gauge theories.

The action in Eq.~(\ref{eq:3_5}) has the same structure as in Eq.~(\ref{eq:3_1}), thus, again, by expanding $e^{-S}$ and collecting the corresponding terms, we can set up a series of equations that we have to solve to obtain the free parameters of the RBF model. The corresponding five equations can be written as follows:
\begin{equation}
F(x_1,x_2,x_3,x_4) = \sum_k a_k \, e^{-A\sum_{j=1}^{4} (x_j-c_{k}^{(j)})^2}
\end{equation}
\begin{eqnarray}
&&F(x_1,x_2,x_3,x_4) \, x_m = \sum_k a_k \, b_{k}^{(m)} \,  e^{-A\sum_{j=1}^{4} (x_j-c_{k}^{(j)})^2}, \nonumber \\
&&\qquad\qquad\qquad\qquad \qquad \text{for} \quad (m=1,2,3,4),
\end{eqnarray}
where we have used the following notation for simplicity:
\begin{equation}
\label{eq:x1234}
x_1=U_{ij}^{(0)}, \quad x_2=U_{ij}^{(1)}, \quad x_3=U_{i-1,j}^{(0)}, \quad x_4=U_{i,j-1}^{(1)}.
\end{equation}

By introducing $C_a$, $C_b^{(1)}$, $C_b^{(2)}$, $C_b^{(3)}$, and $C_b^{(4)}$ fine-tune parameters, and applying the same training method that we described in the previous section, we arrive at the following dependencies between the free parameters:
\begin{equation}
a_k = C_a F(c_k^{(1)},c_k^{(2)},c_k^{(3)},c_k^{(4)}),
\end{equation}
\begin{equation}
a_k\, b_{k}^{(m)} = C_b^{(m)}\, F(c_k^{(1)},c_k^{(2)},c_k^{(3)},c_k^{(4)}) \, c_k^{(m)}, 
\end{equation}
which again comes from the fact that we have fixed the centers through the half-width of the Gaussians, given by the $k_A$ parameter.
The fine-tune parameters are again set so that the overlaps of the Gaussians do not cause too much bias in the intermediate points. 
Using this parametrization, we can set up the corresponding optimization problem as
\begin{equation}
\arg \!\!\! \min_{\!\!\!\!\!\!\!\!\!\! \!\!\! \{ k_A,C_a,C_b^{(m)} \} } \left\{ 
E_1(k_A,C_a) + \sum_{m=1}^4 E_{m+1}(k_A,C_b^{(m)})
\right\}
\end{equation}
where the error terms can be given by
\begin{eqnarray}
&&E_1(k_A,C_a)\!=\! \Bigg\{ \!\frac{1}{N_{T}}\sum_{i=1}^{N_{T}} \Bigg( \!F(x_1,x_2,x_3,x_4)-  \\
&&\sum_{k=1}^K \,C_a \, F(c_k^{(1)},c_k^{(2)},c_k^{(3)},c_k^{(4)}) \, G_k(x_1,x_2,x_3,x_4)  \Bigg)^2\Bigg\}, \nonumber
\end{eqnarray}
and
\begin{eqnarray}
&&E_{m+1}(k_A,C_b^{(m)}) =  \Bigg\{\frac{1}{N_{T}}\sum_{i=1}^{N_{T}} \Bigg(  F(x_1,x_2,x_3,x_4) x_m -  \\
&& \sum_{k=1}^K \,C_b^{(m)}c_k^{(m)}  \, F(c_k^{(1)},c_k^{(2)},c_k^{(3)},c_k^{(4)}) \, G_k(x_1,x_2,x_3,x_4)\Bigg)^2 \Bigg\}, \nonumber
\end{eqnarray}
where we used the following notation for the Gaussian terms:
\begin{equation}
G_k(x_1,x_2,x_3,x_4) = e^{-A\sum_{j=1}^{4} (x_j-c_{k}^{(j)})^2}.
\end{equation}
The optimization procedure might seem complicated at first sight due to the many equations one needs to satisfy, however, setting up the equations is very straightforward, and by using the described training method for the $C_a$, $C_b^{(m)}$ fine-tune parameters, the full optimization procedure only takes a few seconds for a given $k_A$. 

To test the method in 1+1 dimensions let us set up the $F$ function so that it depends only on $U_{i-1,j}^{(0)}$ and $U_{ij}^{(1)}$ as follows:
\begin{equation}
\label{eq:Ex2_F}
F(U_{i-1,j}^{(0)}, U_{ij}^{(1)}) = e^{-2\, U_{i-1,j}^{(0)} U_{ij}^{(1)}}
\end{equation}
While this function does not depend on all the possible external fields that is present in the hopping terms, it is good to show that the method works even for the case when $F$ contains the different combinations of such variables. 

Setting $k_A=20$ ($A \approx 1109$) is generally sufficient for such smooth functions, in which case the fine-tune parameters are $C_a=0.8$ and $C_b^{(m)}=0.8$ for $m=1,2,3,4$. In fact, the fine-tune parameters only have a slight dependency on $k_A$ or even on the applied $F$ function, however in the case of a function with high frequency oscillations a larger $k_A$ might be necessary.
Using the trained parameters the RBF approximation of the of the path integral can be expressed by applying the same factorization formula we have described in the previous section. After collecting the different terms, we get
\begin{equation}
Z^{RBF} =    R^{2N_xN_y}\left(\frac{\pi}{A}\right)^{2N_xN_y} \hat{\mathcal{A}}^{N_xN_y} \det(m,\hat{\mathcal{A}},\hat{\mathcal{B}}_m) ,
\end{equation}
where $N_x$ and $N_y$ are the number of lattice sites in the two directions, and the factor $2$ are coming from the fact that we have to integrate out to $U_{ij}^{(0)}$ and $U_{ij}^{(1)}$ as well.
The $\hat{\mathcal{A}}$, and $\hat{\mathcal{B}}_m$ parameters are given by the combination of the $C_a$, $C_b$ fine-tune parameters and the $c_k$ centers as
\begin{eqnarray}
\label{eq:AB}
&&\hat{\mathcal{A}} = C_a\sum_{k=1}^K  F(c_k^{(2)},c_k^{(3)}),  \\
 &&\hat{\mathcal{B}}_{m}=C_b^{(m)} \sum_{k=1}^K c_k^{(m)} F(c_k^{(2)},c_k^{(3)}), \nonumber
\end{eqnarray}
for $m=1,2,3,4$. 
The fermionic determinant in this case can be expressed as
\begin{equation}
\det(m,\hat{\mathcal{A}},\hat{\mathcal{B}}_m) = \prod_{ij} d\bar{\chi}_{ij}d\chi_{ij} \, e^{-\hat{S}_{ij}^{\mathcal{A},\mathcal{B}_m}[\bar{\chi},\chi]},
\end{equation}
where the fermionic action $\hat{S}_{ij}^{\hat{\mathcal{A}},\hat{\mathcal{B}}_m}[\bar{\chi},\chi]$ is given by
\begin{eqnarray}
\label{eq:35}
&&\hat{S}_{ij}^{\hat{\mathcal{A}},\hat{\mathcal{B}}_m}[\bar{\chi},\chi]  = m\bar{\chi}_{ij}\chi_{ij} + \eta_{ij}^{(0)}  \Bigg[ \frac{1}{2\mathcal{A}} \Big( \hat{\mathcal{B}}_1 \bar{\chi}_{ij}\chi_{i+1,j} -
 \\
&&\hat{\mathcal{B}}_2 \bar{\chi}_{ij}  \chi_{i-1,j}  \Big)  \Bigg] \!\!+\! 
\eta_{ij}^{(1)}  \Bigg[ \frac{1}{2\mathcal{A}} \Big( \hat{\mathcal{B}}_3 \bar{\chi}_{ij} \chi_{i,j+1} - \hat{\mathcal{B}}_4 \bar{\chi}_{ij} \chi_{i,j-1}  \Big)  \Bigg] . \nonumber
\end{eqnarray}

\begin{figure}[!h]
\centering\includegraphics[width=3in]{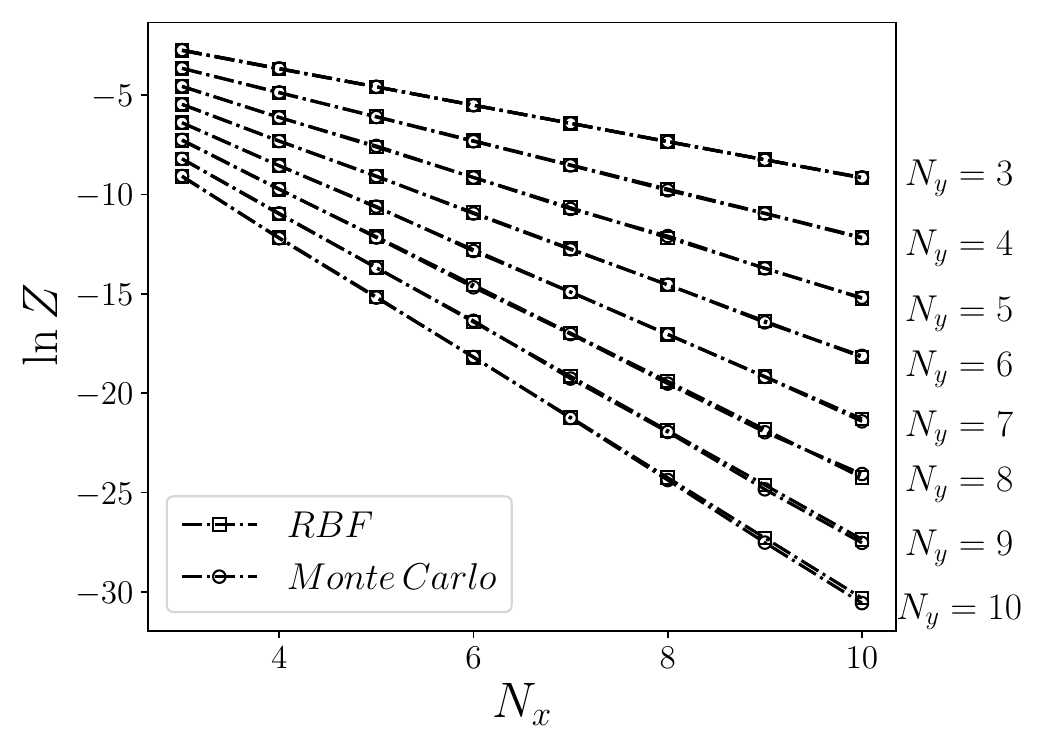}
\caption{Comparison of the Monte Carlo results with the RBF approximations at different lattice sizes $(N_x,N_y)$, for the system defined in Eq.~(\ref{eq:3_3}) with $F(U_{i-1,j}^{(0)},U_{i,j}^{(1)})=e^{-2U_{i-1,j}^{(0)}U_{i,j}^{(0)}}$.
The parameters of the RBF network are set to $k_A=20$, $C_a=0.8$, and $C_b^{(m)}=0.8$ for ($m=1,2,3,4$). The factorization parameter has been fitted to $R\approx 0.995$ through the Monte Carlo point at $(N_x,N_y)=(3,3)$.
}
\label{fig:U1_2}
\end{figure}

Using the RBF approximated path integral the $R$ parameter can be easily fitted by comparing the approximation to the Monte Carlo calculation at some small lattice. This has been done for $(N_x,N_y)=(3,3)$  giving $R\approx 0.995$. The results for different lattice sizes can be seen in Fig.~\ref{fig:U1_2}, showing a very good, around a few percentage accuracy.
The generalization to higher dimensions can be done in the same way, by introducing more variables (more centers) into the Gaussian kernels. 

While approximating the partition function itself could be an effective method to calculate the thermodynamics of finite temperature systems, however in practice many of the interesting observables are calculated as an expectation of specific field insertions e.g. propagators through the insertion of the Grassmann variables, or potentials through Polyakov loops. In the next section we will show that the RBF method can be readily applied to such problems as well.

\section{Field insertions and expectation values}
\label{sec:3}
In this section we will give a method on how to calculate the expectation values of specific field insertions through the RBF approximation. Here, we will only consider the insertion of the external $U(x)$ fields, which in the case of e.g. gauge theories can be used to find phase transition points at finite temperatures, or interaction potentials.

In general (for the systems considered in this paper), the expectation values of specific insertions of the $U(x)$ fields are calculated as a normalized sum of weighted paths given by
\begin{equation}
\label{eq:FI_1}
\langle G(U) \rangle = \frac{\int \mathcal{D}\bar{\chi} \, \mathcal{D}\chi \, \mathcal{D}U \, G(U) \,e^{-S[\bar{\chi},\chi,U]}}{\int \mathcal{D}\bar{\chi} \, \mathcal{D}\chi \, \mathcal{D}U \, e^{-S[\bar{\chi},\chi,U]}
},
\end{equation}
where $G(U)$ is some combination of the $U(x)$ fields at specific space-time points. In the case $U(x)$ is some gauge field e.g. in 1 dimensions the measure could be written as $\mathcal{D}U = \prod_i dU_i F(U_i, U_{i-1},...)$, in which case we get back the form that has been used in the previous sections. While the denominator is just the partition function that has been examined previously, the nominator now also consists of an extra $G(U)$ function that has to be addressed. The main distinction between this $G(U)$ function, and the $F(U)$ function e.g. in Eq.~(\ref{eq:5}) is that while $F(U)$ was given as a product over the full lattice, the $G(U)$ function is a combination of $U_i$'s at a finite number (much smaller than the number of lattice sites) of lattice points.

Here, we will show that by using the RBF expansion, and the factorization formula in Eq.~(\ref{eq:37}), the expectation values can also be approximated at large $N$ by a good accuracy. As the method can be easily extended to higher dimensions, we will do the derivation in one dimension for simplicity.

First, let us denote the denominator in Eq.~(\ref{eq:FI_1}) by $Z_0$, while the nominator by $Z_G$. According to factorization formula $Z_0$ can be expressed as
\begin{equation}
\label{eq:FI_2}
Z_0 \approx R^N \left( \frac{\pi}{A}\right)^N  \int \prod_i d\bar{\chi}_i d\chi_i \Bigg[ \sum_{k=1}^K a_k e^{-S_{k,i}^{RBF}[\bar{\chi},\chi]}\Bigg],
\end{equation}
where the $a_k$ coefficients depend on the $F(U)$ function at the given centers, and $S_{k,i}^{RBF}$ is the fermion action parametrized by the RBF network. The factorization heavily relies on the fact that $Z_0$ is given by a full product over the whole lattice, in which case a stable $R$ parameter can be taken out, and the $U$ dependent part can be separated and integrated out from the determinant. In $Z_G$ the complication arises from the insertion of a finite product of $U_i$ fields, e.g. $G(U)=U_0U_1U_2$, which will perturb the factorization at small lattice sizes. 

To illustrate, let us assume that $G$ is given by
\begin{equation}
\label{eq:FI_3}
G(U) = \prod_{i=1}^{N} U_{i}^{\alpha_i}
\end{equation}
where $\alpha_i$ also acts as an indicator variable, e.g., it is zero if site $i$ does not have an extra field insertion. Putting this back into Eq.~(\ref{eq:22}) we arrive at
\begin{equation}
\label{eq:FI_4}
\!\!\! Z_G = \int \Bigg[  \prod_{i=1}^N d\bar{\chi}_i \, d\chi_i \, dU_i \,
 U_{i}^{\alpha_i} \, F(U_i,U_{i-1}) \, e^{-S_i[\bar{\chi},\chi,U]}  \Bigg],
\end{equation}
where (as before) we have assumed that $F$ only depends on $U_i$, and $U_{i-1}$. In the followings, let us define $N_G>0$ as the number of sites, where $\alpha_i>0$, i.e., where there is an extra field insertion.

Applying the RBF expansion to this system now means that we will have different $a_k$ coefficients at each site, where the approximable functions differ. In this case we will have $N-N_G$ RBF network with the same coefficients, while an additional $N_G$ networks with different $a_k$ coefficients due to the possibly different $\alpha_i$ exponents. This can be represented as
\begin{equation}
\label{eq:FI_6}
F(U_i,U_{i-1})\,e^{-S_i[\overline{\chi},\chi,U]} \approx \sum_{k=1}^{K} a_k^{(0)}e^{S_{k,i}^{RBF}[\overline{\chi},\chi]}e^{-A(U_i-c_k)^2} ,
\end{equation}
for the case, when $\alpha_i=0$ and
\begin{eqnarray}
U_{i}^{\alpha_i} F(U_{i},U_{i-1}) \, e^{-S_{i}[\bar{\chi},\chi,U]} \approx \\
\sum_{k=1}^{K} a_k^{(\alpha_i)}e^{S_{k,i}^{RBF}[\overline{\chi},\chi]}e^{-A(U_{i}-c_k)^2} , \nonumber
\end{eqnarray}
when $\alpha_i>0$. Note that the fermion part of the expansion remains the same, however the $a_k$ coefficients differ in each case where the left side is different. This will change the fermion determinant as well due to the dependency on the $a_k$ coefficients in the final hopping terms through $\sum_k a_k b_k/ \sum_k a_k$.

According to the factorization formula shown e.g., in Eq.~(\ref{eq:FI_2}), each site where neither the site itself nor its $i+1$ neighbor contains any insertions corresponds to the same $R$ parameter
that we would get from $Z_0$. 
On the other hand, at sites, where there are extra insertions, we will get a different $R_i$ coefficient that correspond to the correlations between those and its neighboring sites. This property is further discussed at the end of Appendix A. Using those results, the RBF approximation in this case can be given as follows: 
\begin{eqnarray}
\label{eq:FI_7}
&&Z_G \approx R^{N} \left(\frac{\pi}{A} \right)^N \prod_{i=1}^{\hat{N}_G} f_i \times
  \\
&&
\int \Bigg\{   \prod_{i=1}^N d\bar{\chi}_i \, d\chi_i \sum_{k=1}^K a_k^{(\alpha_i)} e^{-S_{k,i}^{RBF}[\bar{\chi},\chi]}
 \Bigg\}, \nonumber
\end{eqnarray}
where $f_i$ are the ratios of the correlation factors without, and with field insertions defined as $f_i=R_i/R$, while $\hat{N}_G \geq N_G$ is the number of sites, where a correction is needed in those factors. In the case of subsequent field insertions, with open boundary conditions, starting from the first site [e.g. $G(U)=U_1U_2$] $\hat{N}_G=N_G$, however in other, more special cases this needs to be carefully considered. A simple example would be the insertion $G(U) = U_2 U_4$, where due to the fact that the $R_i$ factor depends on the covariance between neighboring sites $(i,i+1)$, the $R$ parameter on site $i=1$ and site $i=3$ also has to be corrected, therefore with $\hat{N}_G=4$ the $R_1$, $R_2$, $R_3$, and $R_4$ parameters will be different than $R$, that has been obtained through calculating $Z_0$.

Later in this section, we will show how the $f_i$ parameters can be calculated using specific examples, but in general, using the notation in Appendix A it can be written as
\begin{equation}
\label{eq:F_cov}
f_i \approx  \frac{\left( 1+\frac{\text{Cov}(X_i^{(\alpha_i)},Y_i)}{E[X_i^{(\alpha_i)}]E[Y_i]}\right)  e^{  \frac{\text{Cov}(Z_i^{(\alpha_i)},Z_{i+1}^{(\alpha_i)})}{E[Z_i^{(\alpha_i)}]E[Z_{i+1}^{(\alpha_i)}]} }}{\left( 1+\frac{\text{Cov}(X_i^{(0)},Y_i)}{E[X_i^{(0)}]E[Y_i]}\right)  e^{  \frac{\text{Cov}(Z_i^{(0)},Z_{i+1}^{(0)})}{E[Z_i^{(0)}]E[Z_{i+1}^{(0)}]} }},
\end{equation}
where $X_i^{(\alpha_i)}$, $Y_i^{(\alpha_i)}$, and $Z_i^{(\alpha_i)}=X_i^{(\alpha_i)}Y_i^{(\alpha_i)}$ are the parts of the full expansion in the $K^N$ sum, and can be given as follows:
\begin{equation}
X_i^{(\alpha_i)} = a_{\langle k \rangle_i}^{(\alpha_i)} \cdot D_{\langle k \rangle_i},
\end{equation}
\begin{equation}
Y_i= e^{-\frac{A}{2}(c_{\langle k \rangle_i}^{(1)} - c_{\langle k \rangle_{i+1}}^{(2)})^2},
\end{equation}
where $D_{\langle k \rangle_i}$ represents the parts containing Grassmann variables, $a_{\langle k \rangle_i}^{(\alpha_i)}$ are the weights of the Gaussian kernels in the RBF expansion, while $c_{\langle k \rangle_i}^{(1)}$, $c_{\langle k \rangle_i}^{(2)}$ are the equidistantly distributed centers. In the case of 'large enough' $A$ (i.e. very narrow Gaussians, and closely placed centers) the centers can be assumed as uniformly distributed random variables, therefore, a function of such variables can be approximated by an integral e.g. $\sum_{\langle k \rangle} F(c_{\langle k \rangle}^{(1)}, c_{\langle k \rangle}^{(2)}) \propto \int dc_{\langle k \rangle}^{(1)} dc_{\langle k \rangle}^{(2)}F(c_{\langle k \rangle}^{(1)}, c_{\langle k \rangle}^{(2)})$. 

 It is also important to note that both $a_{\langle k \rangle_i}^{(\alpha_i)}$ and $D_{\langle k \rangle_i}^{(\alpha_i)}$ depend on the centers $c_{\langle k \rangle_i}^{(1)}$ and $c_{\langle k \rangle_i}^{(2)}$, therefore the covariances between $X_i$ and $Y_i$ could have significant contributions depending on the actual functions involved. 

By applying the same method as before, the fermion parts can be rearranged into having some global $\mathcal{\hat{A}}$ and $\mathcal{\hat{B}}$ parameters. Due to the fact that both $\mathcal{\hat{A}}$ and $\mathcal{\hat{B}}$ depend on the $a_k$ weights the whole expression can be given as

\begin{eqnarray}
&& Z_G \approx R^N \left(\frac{\pi}{A} \right)^N \prod_{i=1}^{\hat{N}_G} f_i \times \\
&& \int \Bigg\{  \prod_{i=1}^N d\bar{\chi}_i \, d\chi_i   
\mathcal{\hat{A}}^{(\alpha_i)} e^{-\hat{S}_{i}^{\hat{A}^{(\alpha_i)}, \hat{B}^{(\alpha_i)}} [\overline{\chi},\chi]} \Bigg\}, \nonumber
\end{eqnarray}
where the RBF fermion action $\hat{S}_i^{ \hat{\mathcal{A}}^{(j)} ,  \hat{\mathcal{B}}^{(j)} }[\overline{\chi},\chi]$ is given by
\begin{eqnarray}
&&\hat{S}_i^{ \hat{\mathcal{A}}^{(\alpha_i)} ,  \hat{\mathcal{B}}^{(\alpha_i)} }[\overline{\chi},\chi] = 
m\overline{\chi}_i\chi_i + \nonumber \\ 
&&\qquad \qquad \frac{\eta_i}{2\hat{\mathcal{A}}^{(\alpha_i)}} \Big(  
\hat{\mathcal{B}}_1^{(\alpha_i)}  \overline{\chi}_i \chi_{i+1} - \hat{\mathcal{B}}_2^{(j)}  \overline{\chi}_i \chi_{i-1}
\Big),
\end{eqnarray}
where the $\hat{\mathcal{A}}^{(\alpha_i)}$ and $\hat{\mathcal{B}}^{(\alpha_i)}$ coefficients can be calculated from the  $a_k^{(\alpha_i)}$ and $b_k^{(\alpha_i)}$ parameters using the corresponding functions at each lattice site as before, with the distinction that now these function could differ from each other due to the field insertions. 

By denoting the fermion determinant as $\det(m,\hat{\mathcal{A}}^{(G)},\hat{\mathcal{B}}_{1,2}^{(G)})$ the final form of $Z_G$ can be written as
\begin{equation}
\label{eq:ZG_RBF}
Z_G  \approx R^N \! \left(\frac{\pi}{A}\right)^{\!\!N} \left( \prod_{i=1}^{\hat{N}_G} f_i \right)
\, \det(m,\hat{\mathcal{A}}^{(G)},\hat{\mathcal{B}}_{1,2}^{(G)})
\prod_{i =1}^{N} \hat{\mathcal{A}}^{(\alpha_i)} .
\end{equation}
The difference between $Z_G$, and $Z_0$ can be summarized as follows: \\
\\
(1) The extra $f_i$ factors, that can be calculated as the ratios of the correlations between the system with, and without field insertions;
\\\\
(2) The different $\hat{\mathcal{A}}^{(\alpha_i)}$ coefficients that are made up by the $a_k^{(\alpha_i)}$ and $b_k^{(\alpha_i)}$ expansion parameters, where $\alpha_i=0$ corresponds to the same values what we get for $Z_0$;
\\\\
(3) The fermion determinant that is made up by the new  $\hat{\mathcal{A}}^{(\alpha_i)}$ and  $\hat{\mathcal{B}}^{(\alpha_i)}$ factors, calculated using $a_k^{(\alpha_i)}$ and $b_k^{(\alpha_i)}$.
\\
\\

To make the method more explicit, let us apply it to a simple example by setting $G(U)$ and the $F(U)$ functions as follows:
\begin{equation}
G(U) = U_1 U_2 U_3,
\end{equation}
\begin{equation}
F(U_i,U_{i-1}) = e^{-\beta U_i U_{i-1}},
\end{equation}
with some $\beta$ parameter. The approximable path integral in this case is given by
\begin{equation}
\label{eq:ZG}
Z_G = \int \prod_i d\overline{\chi}_i d\chi_i dU_i  \Big\{ U_1 U_2 U_3\,e^{-\beta U_i U_{i-1}} e^{-S_i[\overline{\chi},\chi,U]}\Big\},
\end{equation}
with the action
\begin{equation}
S_i[\overline{\chi},\chi,U] = m\overline{\chi}_i\chi_i + \frac{\eta_i}{2} (\overline{\chi}_i U_i \chi_{i+1}
- \overline{\chi}_iU_{i-1}\chi_{i-1} ).
\end{equation}

In practice, we often want to measure the parameter dependency of the probabilities i.e. normalized expectation values. In our case this would mean the dependency of the probability on the $\beta$ parameter, defined through the ratio
\begin{equation}
\label{eq:Q}
P(\beta) = \langle G(U) \rangle_{\beta} = \frac{Z_G(\beta)}{Z_0(\beta)} .
\end{equation}
Such observables could give us insight into the phase transitions of the underlying systems, where the $\beta$ dependency is not trivial. Even though this simple one dimensional example is clearly not enough to study such effects, it is sufficient to show that the RBF model is indeed capable to approximate the ratios with a good accuracy.

To do this, we will first need to approximate the value of $R$ at different $\beta$ parameters, that (as before) can be fitted through calculating $Z_0(\beta,N)$ for small $N$ values. 
In this calculation, we have only used $Z_0(\beta,N=3)$ to fit $R(\beta)$, at $\beta=[0,0.5,1,1.5,2,2.5,3]$.

 The next step is to determine $Z_G$, which requires the recalculation of the $a_k$, and $b_k$ parameters at sites $i=1,2,3$, where we have the same $U_i$ extra field insertions, thus $\alpha_{i=1,2,3}=1$. Using the same techniques as it was shown in Sec. IIC, e.g. in Eqs. (36), (37), (38), and (39), this means the following relations for $a_k^{(\alpha_i)}$, $b_k^{(1)(\alpha_i)}$, and $b_k^{(2)(\alpha_i)}$:
\begin{eqnarray}
\label{eq:akbk}
a_k^{(\alpha_i)}&=&C_a \cdot F(c_k^{(1)},c_k^{(2)})\cdot (c_k^{(1)})^{\alpha_i},\nonumber \\
a_k^{(\alpha_i)} b_k^{(1)(\alpha_i)}&=&C_b^{(1)} c_k^{(1)} \cdot F(c_k^{(1)},c_k^{(2)}) \cdot (c_k^{(1)})^{\alpha_i}, \\
a_k^{(\alpha_i)} b_k^{(2)(\alpha_i)}&=&C_b^{(2)} c_k^{(2)} \cdot F(c_k^{(1)},c_k^{(2)})\cdot (c_k^{(1)})^{\alpha_i}, \nonumber
\end{eqnarray}
where the extra $\cdot (c_k^{(1)})^{\alpha_i}$ factor means the additional $U_i$ insertion at site $i$, with exponent $\alpha_i$. In the case of $i=1,2,3$, we have $\alpha_i=1$, while every other case has $\alpha_i=0$. Also, note that due to the same dimensonality as in the case of $Z_0$ (when all $\alpha_i=0$) the same $C_a$, $C_b^{(1)}$, and $C_b^{(2)}$ fine-tune parameters can be used i.e. $C_a=C_b^{(1)} =C_b^{(2)}=0.89$.  
After, we get the $a_k^{(\alpha_i)}$, $b_k^{(1)(\alpha_i)}$, and $b_k^{(2)(\alpha_i)}$ parameters, the necessary $\hat{\mathcal{A}}^{(\alpha_i)}$ and  $\hat{\mathcal{B}}^{(\alpha_i)}$ factors, and $\det(m,\hat{\mathcal{A}}^{(G)},\hat{\mathcal{B}}_{1,2}^{(G)})$ can be calculated.

The final step is to determine the $f_i$ factors from the covariance of the neighboring sites using Eq.~(\ref{eq:F_cov}). First, for simplicity, let us omit the determinant, and any Grassmann variables in the action, by setting $S_i[\overline{\chi},\chi,U]=0$. In this case, we do not have to care about the fermion determinants in $Z_0$ and in $Z_G$, nor the $b_k$ parameters in Eq.~(\ref{eq:akbk}) (second and third equations). By comparing the probabilites with, and without fermions it is possible to show the difference the fermions make, which is also an important aspect of field theories. 

Using these simplifications, let us define the following variables:
\begin{equation}
X_{i}^{(1)} = C_a  \cdot e^{-\beta c_i^{(1)}c_i^{(2)}} \cdot c_i^{(1)}
\end{equation}
\begin{equation}
X_{i}^{(0)} = C_a  \cdot e^{-\beta c_i^{(1)}c_i^{(2)}} 
\end{equation}
\begin{equation}
Y_{i} =  e^{-\frac{A}{2}(c^{(1)}_i - c^{(2)}_{i+1})^2},
\end{equation}
where we have omitted the $k$ indices from the centers, because due to the large $A$ limit, we assume that each center is a uniformly distributed random variable in $c_i \in[0,1]$. From $X_{i}^{(1)}$, $X_{i}^{(0)}$, and $Y_i$. To calculate $f_i$ we will need to determine the following expectation values and covariances:
\begin{eqnarray}
&& E [X_{i}^{(1)}], \quad E[X_i^{(0)}], \quad E[Y_i], \quad E[X_{i}^{(1)},Y_i], \nonumber \\
 &&E[X_{i}^{(0)},Y_i], \quad \text{Cov}[X_{i}^{(1)},Y_i], \quad  \text{Cov}[X_{i}^{(0)},Y_i],  \\
 && \text{Cov}[X_{i}^{(1)}Y_i, X_{i+1}^{(1)}Y_{i+1}], \quad  \text{Cov}[X_{i}^{(1)}Y_i, X_{i+1}^{(0)}Y_{i+1}], \nonumber
\end{eqnarray}
where each expectation value (and covariances) are represented by an integral to the center variables. The last line represents to covariances between the subsequent sites, where $\text{Cov}[X_{i}^{(1)}Y_i, X_{i+1}^{(1)}Y_{i+1}]$ needs to be calculated for $f_1$, and $f_2$ with neighbors $(1,2)$, and $(2,3)$, while $\text{Cov}[X_{i}^{(1)}Y_i, X_{i+1}^{(0)}Y_{i+1}]$ corresponds to the neighbors $(3,4)$ i.e., $f_3$. 

\begin{figure}[!h]
\centering\includegraphics[width=3.2in]{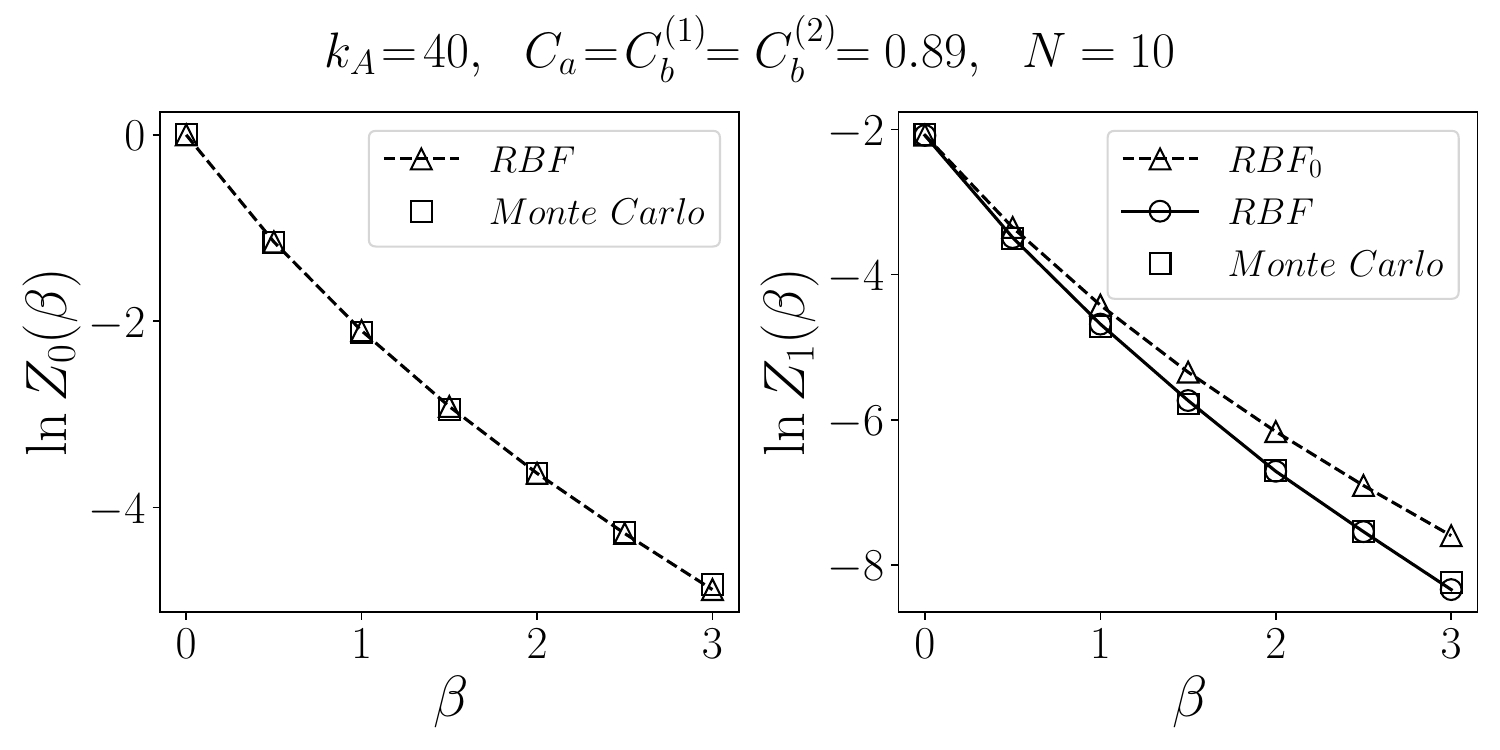}
\caption{Comparison of $\ln Z_0(\beta)$ and $\ln Z_G(\beta)$ at $N=10$, with the Monte Carlo results, using an RBF network with $k_A=40$ kernels for the system without fermions. The right-hand figure also shows the case, when all the $f_i$ coefficients are set to $1$ ($RBF_0$), while ($RBF$) shows the full calculations for $Z_G(\beta)$.
}
\label{fig:PB_0}
\end{figure}

After calculating the integrals, the $f_1$, $f_2$, and $f_3$ coefficients can be put together using Eq.~(\ref{eq:F_cov}), and we will have all the ingredients to calculate $Z_G$ for the case without fermions. In Fig.~\ref{fig:PB_0}, we show the results for $\ln Z_0(\beta)$ and $\ln Z_G(\beta)$ at $N=10$, compared to the Monte Carlo results. The RBF network is parametrized with $k_A=40$ and $C_a=C_b^{(1)}=C_b^{(2)}=0.89$. For comparison, we also show the case when $f_1$, $f_2$, and $f_3$ are all set to $f_i=1$ (labeled as $RBF_0$), which shows that these corrections are indeed needed, especially with larger $\beta$ values.

\begin{figure}[!h]
\centering\includegraphics[width=3in]{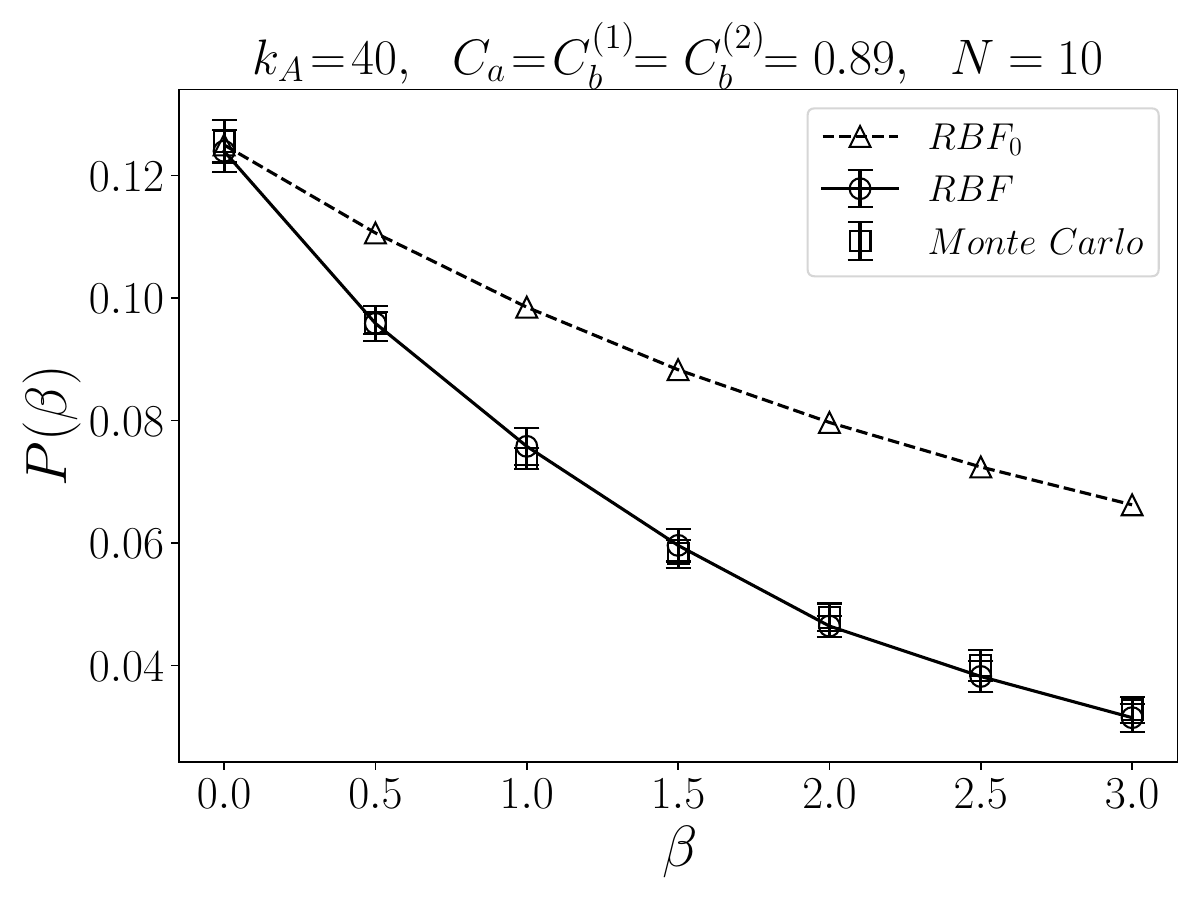}
\caption{Comparison of the probabilities between the RBF approximation and the Monte Carlo results at different $\beta$ values for the system without fermions. ($RBF_0$) shows the results with all $f_i$'s set to $1$, while ($RBF$) is calculated using the full network.
}
\label{fig:PB_1}
\end{figure}

After, we have an approximation for $Z_0(\beta)$ and $Z_G(\beta)$ at some specific $N$, the probabilities can be readily calculated by taking their ratios. This is shown in Fig.~\ref{fig:PB_1}, where a very good match is achieved between the RBF approximation and the Monte Carlo results.

\begin{figure}[!h]
\centering\includegraphics[width=3in]{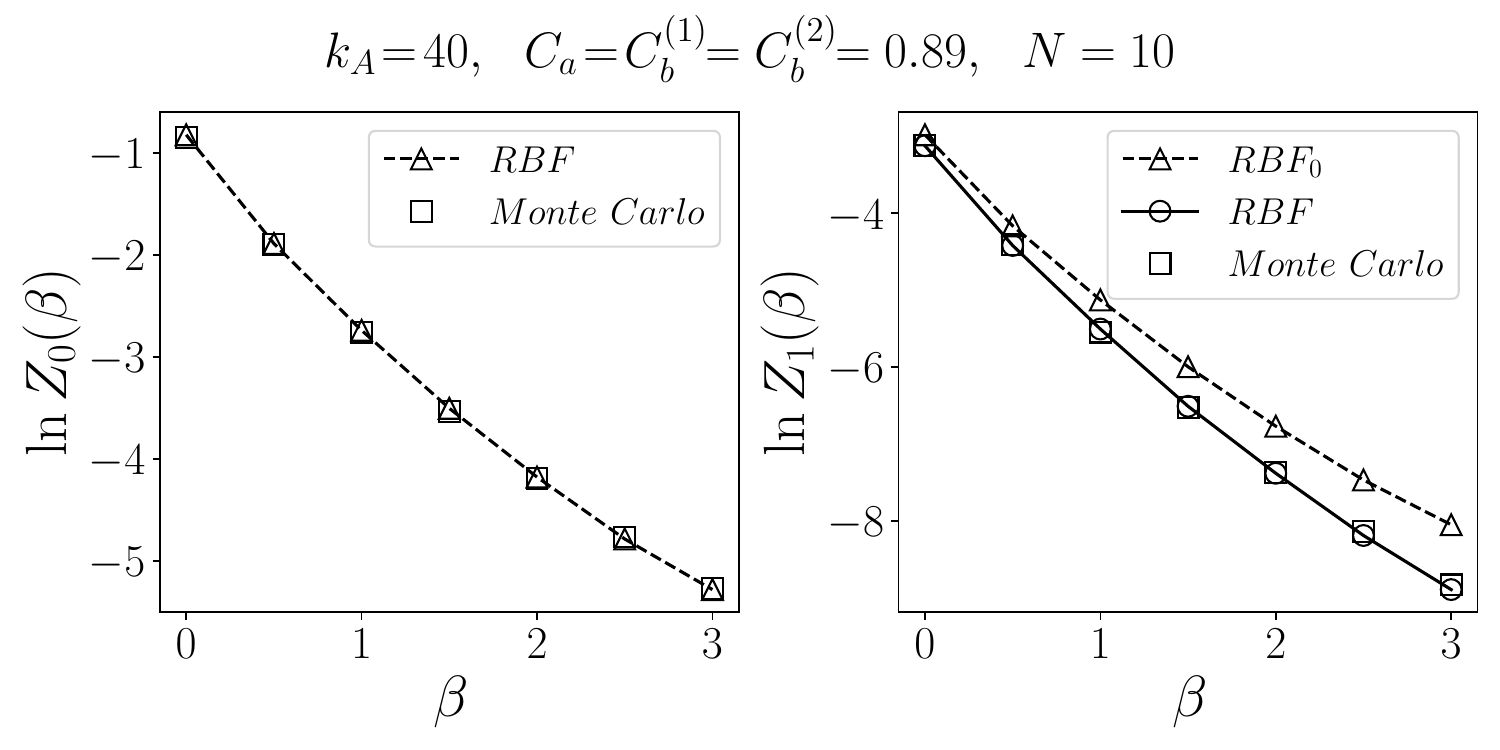}
\caption{Comparison of $\ln Z_0(\beta)$, and $\ln Z_G(\beta)$ at $N=10$, with the Monte Carlo results, using an RBF network with $k_A=40$ kernels for the system with fermions included. The right-hand figure also shows the case, when all the $f_i$ coefficients are set to $1$ ($RBF_0$), while ($RBF$) shows the full calculations for $Z_G(\beta)$.
}
\label{fig:PB_20}
\end{figure}

\begin{figure}[!h]
\centering\includegraphics[width=3in]{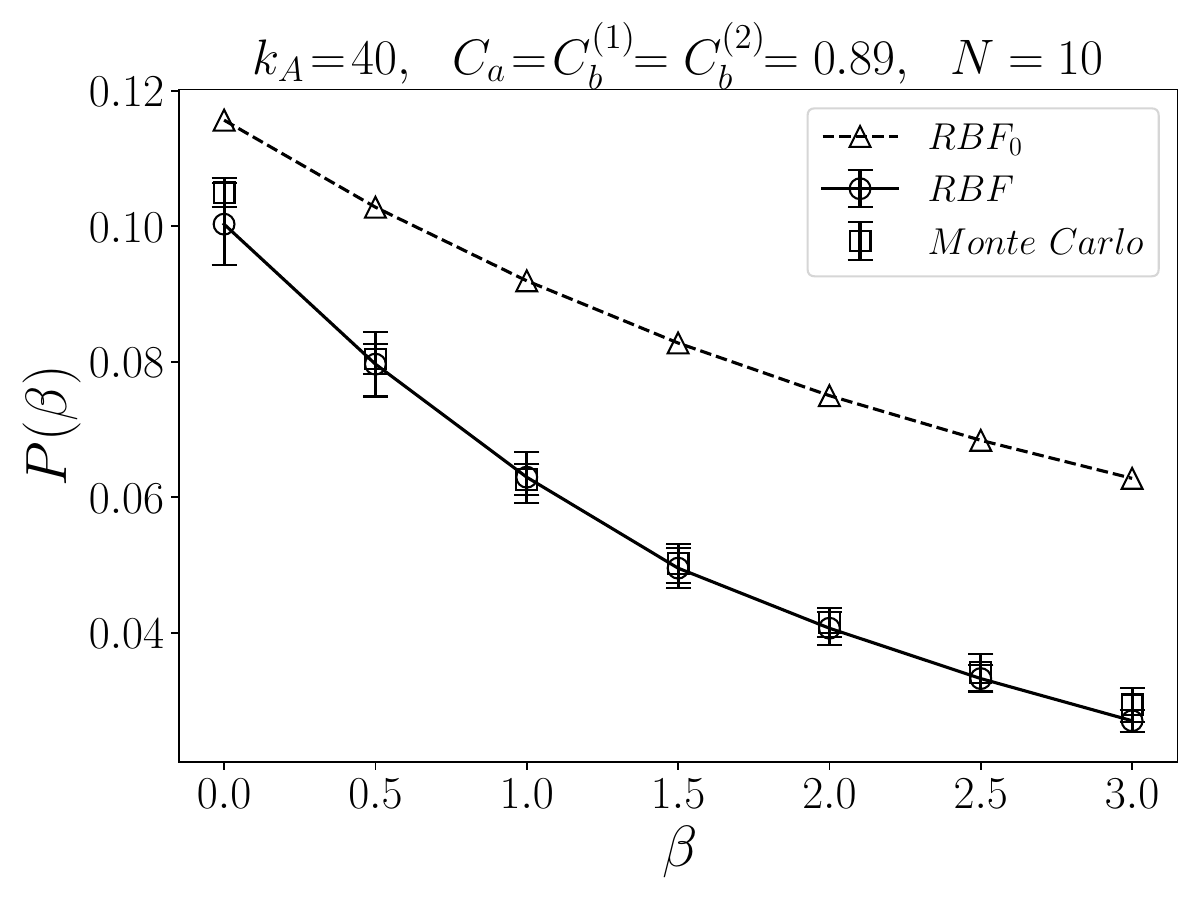}
\caption{Comparison of the probabilities between the RBF approximation and the Monte Carlo results at different $\beta$ values for the system with fermions included. ($RBF_0$) shows the results with all $f_i$'s set to $1$, while ($RBF$) is calculated using the full network.
}
\label{fig:PB_21}
\end{figure}

In the case, when we also include fermions, the situation slightly changes due to the fact that now we need to consider all the equations for $b_k$'s as well in Eq.~(\ref{eq:akbk}), and not just for the $a_k$ parameters. Furthermore, the calculation of $f_i$'s also changes due to additional $D_{\langle k \rangle_i}^{(\alpha_i)}$ determinant parts in $X_i^{\alpha_i}$. In this case we define the following $X_i$ and $Y_i$ variables:
\begin{equation}
X_{i}^{(1)} = C_a  \cdot e^{-\beta c_i^{(1)}c_i^{(2)}} \cdot c_i^{(1)} \cdot D_i(c_i^{(1)},c_i^{(2)})
\end{equation}
\begin{equation}
X_{i}^{(0)} = C_a  \cdot e^{-\beta c_i^{(1)}c_i^{(2)}} \cdot D_i(c_i^{(1)},c_i^{(2)})
\end{equation}
\begin{equation}
Y_{i} =  e^{-\frac{A}{2}(c^{(1)}_i - c^{(2)}_{i+1})^2},
\end{equation}
with
\begin{equation}
D_i(c_i^{(1)},c_i^{(2)}) = m \overline{\chi}_i \chi_i +\frac{\eta_i}{2}\Big( c_i^{(1)}\overline{\chi}_i \chi_{i+1} -  c_i^{(2)}\overline{\chi}_i \chi_{i-1}  \Big),
\end{equation}
where as before the $\langle k \rangle_i$ indices were omitted because each center is assumed as a uniformly distributed random variable. Using the new $X$ variables, the expectation values, and covariances have to be recalculated to be able to give the new $f_i$ ratios. The fermion part will give some extra contributions, but in general the corresponding integrals can be calculated the same way as before. The results for $Z_0(\beta)$ and $Z_G(\beta)$ can be followed in Fig.~\ref{fig:PB_20}, while the probabilities $P(\beta)$ are shown in Fig.~\ref{fig:PB_21}.

In each case, the comparison with the Monte Carlo results shows the same accuracy as what we have seen before, when we neglected the fermions. It can also be seen that including fermions lowers the probabilties, which behavior is well-described by the RBF method for all $\beta$ values. 

\begin{figure}[!h]
\centering\includegraphics[width=3in]{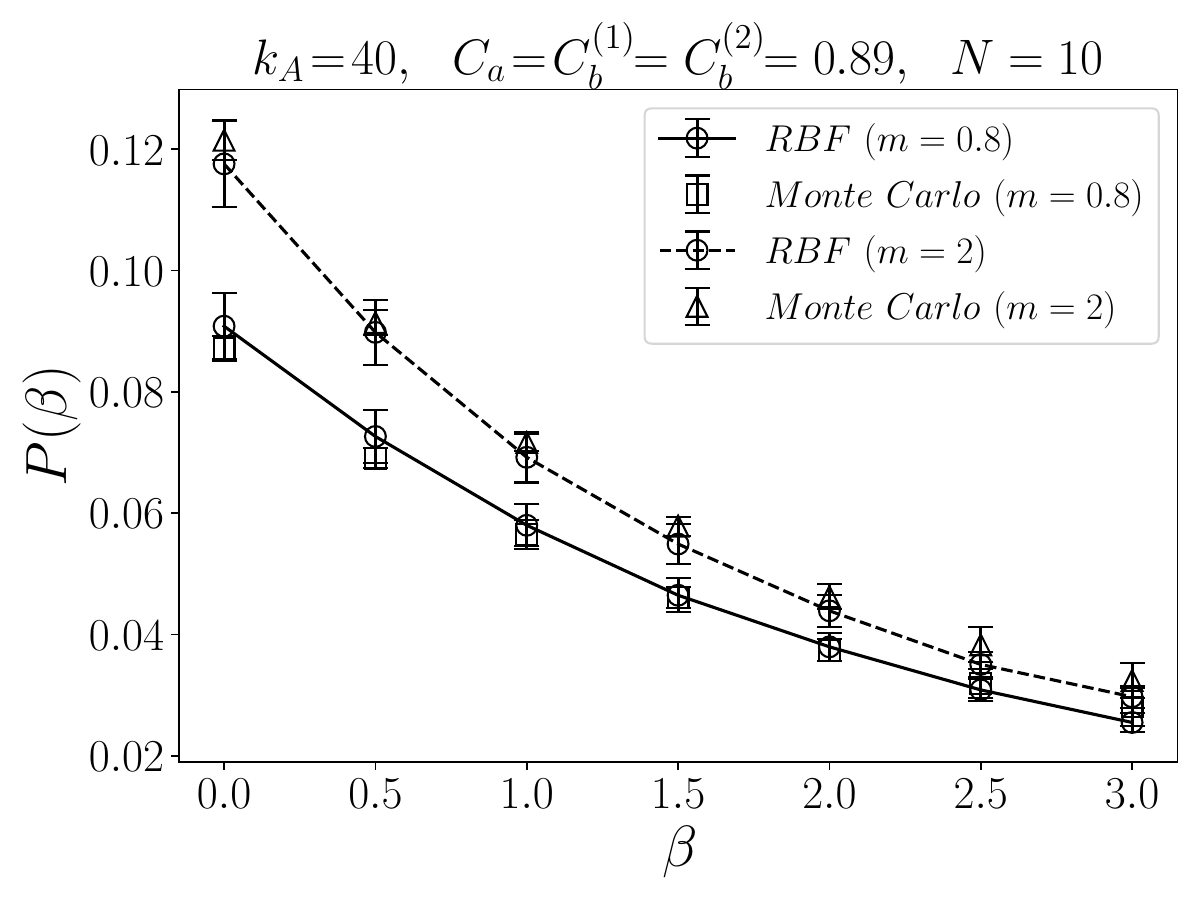}
\caption{Probabilities of the fermion case calculated using $m=0.8$ (full) and $m=2$ (dashed) mass parameters compared to the Monte Carlo results.
}
\label{fig:PB_M}
\end{figure}

As a final example in this section let us study the mass parameter dependency in the case when we still include fermions. As the covariances between sites directly depends on the mass parameter it is expected to change the probabilities in a nontrivial way. To see this we have done the calculations using $m=0.8$ and $m=2$ which can be followed in Fig.~\ref{fig:PB_M}. 

The results show that the probability tends to be larger for larger masses, while the difference between them shrinks as we increase the $\beta$ parameter. The RBF approximation gives back this tendency with a very good, few percentage accuracy compared to the Monte Carlo calculations.
In every case, the errors of the RBF approximations are calculated from the uncertainties of the numerical determination of $f_i$'s (expectation values, and covariances), and the uncertainties of the $R$ factor, that is determined through calculating the $Z_0$ integral at $N=3$.

\section{Additional comments and extensions}
\label{sec:4}
Before we conclude, let us briefly give some final remarks on the RBF approximation and its capabilities. Firstly, in all examples it was assumed that $F$ does not depend on any other variables than $U_i$, and $U_{i-1}$. This assumption is not a restriction, but only used to simplify the explanation of the models. In more complex cases $F$ could also depend on other neighbors of $U_i$ e.g. in SU(2) gauge theory in more than one dimensions, in which case the Polyakov loops would also contain forward neighbors. 

In the case of a functional form of $F=F(U_i,U_{i-1},U_{i+1})$, we will have an extra dimension that is related to the variable $U_{i+1}$, which means we will also have another set of centers $c^{(3)}$. The corresponding RBF network will have the form of
\begin{eqnarray}
 &&F(U_i,U_{i-1},U_{i+1}) \, e^{-S_i[\overline{\chi},\chi,U]} \approx \sum_{k=1}^K a_k \, e^{-S_{k,i}^{RBF}[\overline{\chi},\chi]} \times \nonumber \\
&& \qquad e^{-A \big[ (U_i-c_k^{(1)})^2 + (U_{i-1}-c_k^{(2)})^2 + (U_{i+1}-c_k^{(3)})^2  \big]},
\end{eqnarray}
where $S_{k,i}$ and $S_{k,i}^{RBF}$ both depend on $U_i$ and $U_{i-1}$, and according to the training procedure we have used before $K=k_A^3$, with $k_A$ representing the number of centers in each dimension. This would also mean that in this case $a_k=a_k(c_k^{(1)},c_k^{(2)},c_k^{(3)})$. In the case of $F(U_i,U_{i-1})$, expanding the product of sums into a $K^N$ sum of products, then rearranging the exponents, resulted in $ \int dU_i \, e^{-A [ (U_i-c_{\langle k \rangle_i}^{(1)})^2  + (U_i-c_{\langle k \rangle_{i+1}}^{(2)})^2] }$, which after integrated out was proportional to $ e^{-\frac{A}{2} (c_{\langle k \rangle_i}^{(1)} - c_{\langle k \rangle_{i+1}}^{(2)})^2}$. In the case of $F(U_i,U_{i-1},U_{i+1})$ this will change to
\begin{eqnarray}
&&\int dU_i \, e^{-A \big[(U_i-c_{\langle k \rangle_i}^{(1)})^2  + (U_i-c_{\langle k \rangle_{i+1}}^{(2)})^2 + (U_i-c_{\langle k \rangle_{i-1}}^{(3)})^2\big]  } \propto \nonumber \\
&&
e^{-\frac{A}{3} \big[ (c_{\langle k \rangle_i}^{(1)} - c_{\langle k \rangle_{i+1}}^{(2)})^2 
+(c_{\langle k \rangle_{i+1}}^{(2)} - c_{\langle k \rangle_{i-1}}^{(3)})^2 
+(c_{\langle k \rangle_{i-1}}^{(3)} - c_{\langle k \rangle_{i}}^{(1)})^2 \big]},
\nonumber \\
\end{eqnarray}
with the proportionality factor of $\sqrt{\frac{\pi}{3A}}$. The factorization using this form is also expected to work as before due to the local couplings, however after integrating out to $c_{\langle k \rangle_i}^{(1)}$, $c_{\langle k \rangle_i}^{(2)}$, and $c_{\langle k \rangle_i}^{(3)}$ in the large $A$-limit, we will arrive at a slightly different prefactor, and the path integral in this case will be given as follows:
\begin{equation}
Z \approx R^N \left( \frac{\pi}{A} \right)^{\frac{3N}{2}} \hat{\mathcal{A}}^N \det(m, \hat{\mathcal{A}},\hat{\mathcal{B}}_1,\hat{\mathcal{B}}_2).
\end{equation}
In the case of other types of functional forms, the RBF expansion has to be set up accordingly.

The second remark is about the used discretizaton of the fermions. Here, staggered fermions has been used due to its simplicity to be able to show how the method works. Though this discretization scheme is still used in lattice calculations its continuum properties are not trivial (e.g. rooting trick, symmetries etc.). It turns out that the RBF expansion is not restricted to staggered fermions, but also works for Dirac fermions and possibly other types of local discretization schemes as well. Though this has not been studied in this paper it is a necessary future step that has to be addressed later on, especially if we want to apply the technique to QCD.

Thirdly, due to the fact that the factorization does not depend wether we include fermions or not, the method could also be readily applied to e.g. Yang-Mills theories as well. Calculating probabilities is especially easy due to the structure of the integrals that have to be solved either analytically (in some cases), or numerically to get the necessary covariances and expectation values.

\section{Conclusions}
In this paper we have applied a radial basis function neural network expansion to Euclidean path integrals including fermions as Grassmann variables. The method relies on separating the fermion parts from the dynamics of the external fields through the help of a factorization formula that relies on the applied Gaussian form, and the structure of the RBF expansion. By tuning the parameters such as widths, centers, and weights of the Gaussians, it is possible to estimate the logarithm of the partition function below a percent relative accuracy in seconds even for large lattice sizes.

In the case of staggered fermions, the expansion leads to a set of equations for each term that appears when expanding the exponential term in the path integral. By applying a specific training method, the task of training the RBF model boils down to optimizing for the number of centers, and some corresponding fine-tuning parameters. 
As the dimension grows, the number of equations increases, however, due to the specific structure of the equations, the optimization problem is still relatively easy to solve.
By factorizing the large sum resulted by applying the RBF model, the path integral can be approximated in a very efficient manner in a closed form, using an extra fittable parameter ($R$), that is related to the covariance between the different terms in the expansion. 

Each example that is shown in this paper is used to build up the model and to show its capabilities without additional numerical complications. The extensions to higher dimensions are also discussed. One of the main advantages of the RBF model, is the possibility to approximate expectation values in a very efficient manner using the factorization properties of the radial basis function expansion.
The method is also not exclusive to systems with fermions, and can be easily adjusted to e.g., Yang-Mills theories in which case the fermion determinant should be set to unity. The factorization will have the same properties, and the expectations values can be calculated the same way. 

Preliminary calculations also show that the factorization works even for Dirac fermions, and also with the inclusion of chemical potentials in the temporal directions. This would mean that there is a possibility that the method could also be useful in describing problems at finite densities. In future works we aim to address such problems.

\section*{APPENDIX A - Factorization of the RBF expansion}
One of the main ingredients of the RBF method that has been shown in this work is the factorization formula used e.g. in Eq.~(\ref{eq:36}). In this section we will show through numerical examples that this factorization holds, even for the case when the $F$ function is more general.

In the case of staggered fermions in one dimension, and with a $F=F(U_i,U_{i-1})$ functional form the factorization formula is given by
\begin{eqnarray}
\label{eq:A1}
&&K^N \sum_{\langle k \rangle} \prod_i a_{\langle k \rangle_i} D_{\langle k \rangle_i} e^{-\frac{A}{2} (c^{(1)}_{\langle k \rangle_i} - c^{(2)}_{\langle k \rangle_{i+1}})^2} \approx  \\
&&R^N \Bigg[\sum_{\langle k \rangle } \prod_i a_{\langle k \rangle_i} D_{\langle k \rangle_i} \Bigg]
\Bigg[ \sum_{\langle k \rangle } \prod_i
e^{-\frac{A}{2} (c^{(1)}_{\langle k \rangle_i} - c^{(2)}_{\langle k \rangle_{i+1}})^2}  \Bigg],
\nonumber
\end{eqnarray}
where $D=D(c_{\langle k \rangle_i}^{(1)},c_{\langle k \rangle_i}^{(2)})$ represents the fermion part including the Grassmann integrals [i.e. $\prod_i D_{\langle k \rangle_i} = \det(D_{\langle k \rangle})$], with $c^{(1)}$ and $c^{(2)}$ centers corresponding to $U_{i}$, and $U_{i-1}$ fields, while $a$'s are the RBF weights corresponding to the $F$ function given at some corresponding centers e.g., $a_{\langle k \rangle_i} \sim F(c^{(1)}_{\langle k \rangle_i}, c^{(2)}_{\langle k \rangle_i} )$. The sum goes over all the possible $K^N$ combinations, where $K$ is the number of centers distributed in a two-dimensional grid spanned by $(c^{(1)},c^{(2)})$. The $K^N$ factor in the first line represents the fact that the factorization corresponds to the means of the different combinations i.e., $\frac{1}{K^N} \sum_{\langle k \rangle}$, therefore without simplification the second line would have a factor $\frac{1}{K^{2N}}$. The $A$ width parameter depends on the number of centers through $A=4k_A^2\ln(2)$, where $k_A$ is the number of centers in one dimensions, therefore $K=k_A^2$ in the case of two centers $c^{(1)}$ and $c^{(2)}$.

In its root the problem boils down to the factorization of the expectation values of the product of functions of uniformly distributed random variables (centers), that can be written as
\begin{equation}
\label{eq:EXY1}
E \Bigg[ \prod_i X_i Y_i \Bigg] \approx  \prod_i E\Big[ X_i \Big]  \prod_i E \Big[  Y_i\Big] \prod_i R_i ,
\end{equation}
where $X_i$ represents the first bracket in the second line of Eq.~(\ref{eq:A1}), $Y_i$ represents the second bracketed terms (i.e. the Gaussian factors), while in general $R_i$ depends on the covariances between the variables. First, let us denote $Z_i = X_i Y_i$ and apply a cumulant expansion to $\ln E[\prod_i Z_i]$, in which case we will have the following form after exponentization:
\begin{equation}
\label{eq:Z1}
E \left[ \prod_i Z_i \right] \approx \prod_i E \left[ Z_i \right]  \exp\left( \sum_{\substack{i,j \\ (i<j)}} \frac{\text{Cov}(Z_i,Z_j)}{E[Z_i]E[Z_j]} + \cdots \right),
\end{equation}
where ($\cdots$) means higher-order cumulants that are suppressed due to the Gaussian factors. In the next step let us express $E[Z_i]$ with $X_i$, $Y_i$, and their covariance as
\begin{equation}
E[Z_i]=E[X_iY_i] = E[X_i]E[Y_i] \left(  1 + \frac{\text{Cov}(X_i,Y_i)}{E[X_i]E[Y_i]}\right).
\end{equation}
By putting this form back into Eq.~(\ref{eq:Z1}), and using the fact that $\prod_i E[X_i]=E[\prod_i X_i]$ and $\prod_i E[Y_i]=E[\prod_i Y_i]$, we will arrive at
\begin{eqnarray}
\label{eq:EXY2}
&&E\Bigg[ \prod_i X_i Y_i \Bigg] \approx  \prod_iE\Big[ X_i\Big] \prod_iE\Big[Y_i\Big] \times \\
&&\left[ \prod_i\left( 1+\frac{\text{Cov}(X_i,Y_i)}{E[X_i]E[Y_i]}\right)  \right]  \exp\left( \sum_{\substack{i,j \\ (i<j)}} \frac{\text{Cov}(Z_i,Z_j)}{E[Z_i]E[Z_j]} + \cdots \right) \nonumber
\end{eqnarray}
The narrow Gaussians factors (i.e. $Y_i$) make sure that the dominant contributions of the final term come from the covariance of the nearest neighbors, and the higher-order cumulants are suppressed, therefore the expression in Eq.~(\ref{eq:EXY2}) can be given by the form shown in Eq.~(\ref{eq:EXY1}) with the following $R$ coefficients,
\begin{equation}
R_i \approx \left( 1+\frac{\text{Cov}(X_i,Y_i)}{E[X_i]E[Y_i]}\right)  \exp\left( \frac{1}{N}\sum_{\substack{i,j \\ (i<j)}} \frac{\text{Cov}(Z_i,Z_j)}{E[Z_i]E[Z_j]}  \right) ,
\end{equation}
where we have neglected the higher-order correlations.
In the case, when the partition function is written as a 'full-product' of the same functional form at each site, we arrive at the structure shown in Eq.~(\ref{eq:A1}) with a factorization coefficient that is expected to stabilize to a constant value after some finite lattice size $N$, approximated by
\begin{equation}
R \rightarrow \left( 1+\frac{\text{Cov}(X_i,Y_i)}{E[X_i]E[Y_i]}\right)  \exp\left( \frac{\text{Cov}(Z_i,Z_{i+1})}{E[Z_i]E[Z_{i+1}]}  \right) ,
\end{equation}
where we have kept the $i$ dependence (despite the fact that the covariances and expectation values are the same for every $i$), due to the different possible boundary conditions that could slightly alter the actual $R$ value.
The convergence speed and stability at small $N$ depend on the functions involved, the boundary conditions, etc.

The factorization would be exact with $R\approx1$ if the covariance between the two bracketed terms were negligible compared to their mean values, however this cannot be exactly satisfied due to the fact that both the determinant, and the RBF weights depend on the centers. 

\begin{figure}[!h]
\centering\includegraphics[width=3in]{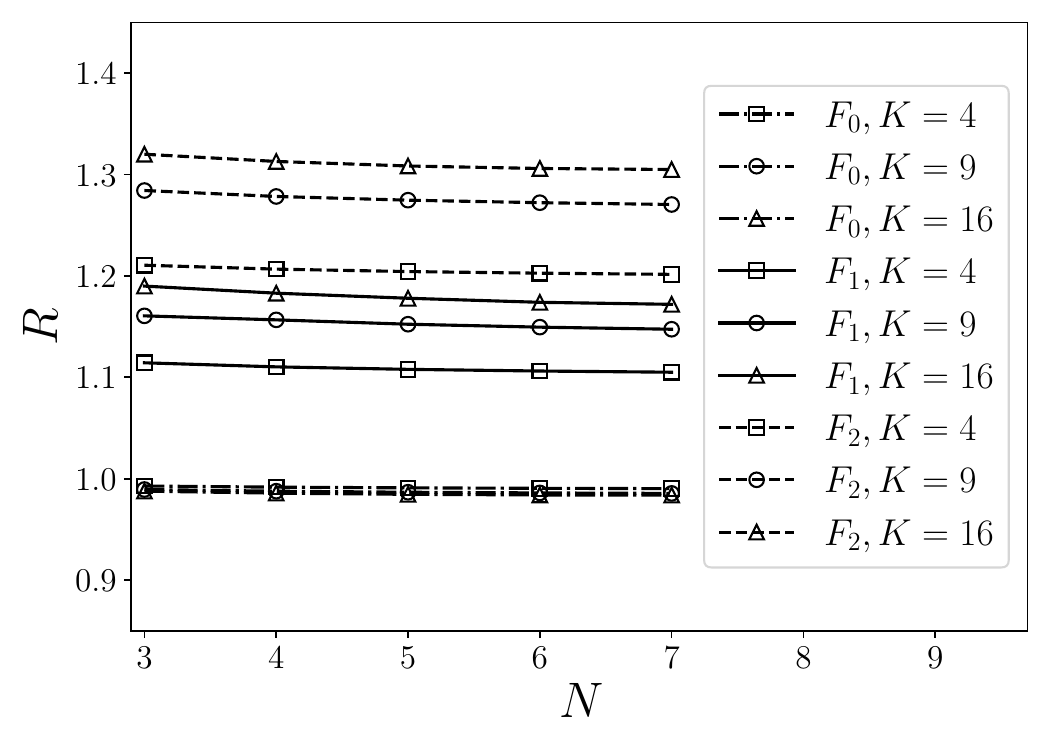}
\caption{ Lattice size ($N$) dependence of the factorization parameter ($R$) for three different $F$ functions given by Eq.~(\ref{eq:F123}), using three different number of centers ($K$) in the RBF networks.
}
\label{fig:A1}
\end{figure}

To show that the factorization indeed works in a controlled environment where we calculate the full $K^N$ sum, we have set the mass parameter to $m=1$, and defined the following three $F(c_{\langle k \rangle}^{(1)}, F(c_{\langle k \rangle}^{(2)})$ functions:
\begin{eqnarray}
\label{eq:F123}
&&F_1 = 1 ,  \nonumber \\
&&F_2 =  c^{(2)}_{\langle k \rangle_i} \, e^{2\, c^{(1)}_{\langle k \rangle_i}c^{(2)}_{\langle k \rangle_i}}, \nonumber \\
&&F_3 = c^{(1)}_{\langle k \rangle_i} \, \sin(c^{(2)}_{\langle k \rangle_i})^2,  
\end{eqnarray}
where $F_1$ is a reference case when only the determinant is coupled to the Gaussians. In Fig.~\ref{fig:A1} we have show the $N$ dependence of the obtained $R$ parameters for $K=4$, $9$, $16$, which corresponds to $k_A=2$, $3$, and $4$.

From the results it can be seen that $R$ is rather stable after $N>3$ in each case, and has a monotone dependency on the number of centers $K$. 
The simplest case with $F_1=1$ the factorization is almost perfect with $R\approx 1$, which means that the covariances are negligible. For more complex functions involving the joint variables, the covariance is larger, thus the larger $R$ values. The monotone increase in $R$ is coming from the fact that in the case of more centers, we are able to sample the corresponding function with a finer resolution, which after some corresponding $K$ will saturate to a finite $R$.

The factorization with a constant $R$ parameter (at 'large' $N$) is possible due to the full product structure of the path integral. In the case of field insertions at specific sites, the factorization formula needs extensions that also considers the 'perturbations' at those specific lattice sites. To illustrate this with a simple example, let us assume that we have an extra contribution at site $i=1$. Due to the fact that in this case the corresponding $a_{\langle k \rangle_1}$ parameters in Eq.~(\ref{eq:A1}) will differ from $a_{\langle k \rangle_{i>1}}$, the corresponding $X_1$ and $X_{i>1}$ factors in [e.g., in Eq.~(\ref{eq:EXY1})] will also differ. To be able to clearly separate the contributions, let us denote the new $X_1$ and $Z_1$ variable by $\tilde{X}_1$, $\tilde{Z}_1$, with the unchanged variables staying at $X_i$, and $Z_i$. Note, that even with the field insertion the Gaussian weight factors $Y_i$ will not change, thus $\tilde{Z}_1=\tilde{X}_1Y_1$. 

This addition will mean that $\prod_i R_i$ in Eq.~(\ref{eq:EXY1}) can be written as
\begin{equation}
\prod_i R_i = \tilde{R}_1 \cdot  R^{N-1} = f_1 \cdot R^N,
\end{equation}
where we have defined $f_1$ as the ratio of the factorization parameters given as
\begin{equation}
f_1 = \frac{\tilde{R}_1}{R_1} = \frac{ \left( 1+\frac{\text{Cov}(\tilde{X}_1,Y_1)}{E[\tilde{X}_1]E[Y_1]}\right)  \exp\left( \frac{\text{Cov}(\tilde{Z}_1,Z_{2})}{E[\tilde{Z}_1]E[Z_{2}]}  \right)}{\left( 1+\frac{\text{Cov}(X_1,Y_1)}{E[X_1]E[Y_1]}\right)  \exp\left( \frac{\text{Cov}(Z_1,Z_{2})}{E[Z_1]E[Z_{2}]}  \right)},
\end{equation}
where $\tilde{X}_1$ and $\tilde{Z}_1$ represents the new variables including the extra field insertions. While it is clear that at large $N$ values, the dominant contribution comes from $R$ (at least for small number of field insertions), in the case when we aim to calculate expectation values i.e., probabilities, by taking the ratios, the additional factors $f_i$'s will be very important. This behavior is clearly seen in Sec.~\ref{sec:3}, where the calculated probabilities are compared with and without including the $f_i$ factors.

\begin{acknowledgments}
This work was supported by the Korea National Research Foundation under Grant No. 2023R1A2C300302311 and 2023K2A9A1A0609492411, and the Hungarian OTKA fund K138277.
\end{acknowledgments}

\bibliography{Main.bib}

\end{document}